\documentclass[superscriptaddress,twocolumn, aps,nofootnoteinbib]{revtex4}

\DeclareRobustCommand{\trace}{%
  \begingroup\normalfont
  \includegraphics[height=\fontcharht\font`\B]{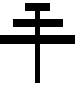}%
  \endgroup
}

\usepackage{bbm}
	
\usepackage{amsmath}
\usepackage{hyperref}

\usepackage{tikz}
\pgfdeclarelayer{nodelayer}
\pgfdeclarelayer{edgelayer}
\pgfsetlayers{edgelayer,nodelayer,main}
\tikzstyle{none}=[inner sep=0pt]
\tikzstyle{new}=[inner sep=2pt]
\usetikzlibrary{shapes, arrows, shadows}
\usetikzlibrary{circuits.ee.IEC}
\usetikzlibrary{decorations.pathmorphing}
\usetikzlibrary{shapes.geometric}

\tikzstyle{env}=[copoint,regular polygon rotate=0,minimum width=0.2cm, fill=black]

\tikzstyle{probs}=[shape=semicircle,fill=white,draw=black,shape border rotate=180,minimum width=1.2cm]

\usetikzlibrary{circuits.ee.IEC}
\usetikzlibrary{decorations.pathmorphing}
\usetikzlibrary{shapes.geometric}

\tikzstyle{wavy}=[decorate,decoration={snake, segment length=1mm, amplitude=0.3mm}]
\tikzstyle{mopoint}=[shape=semicircle, fill=white,draw=black,shape border rotate=180,scale =0.75]
\tikzstyle{mocopoint}=[shape=semicircle, fill=white,draw=black,minimum width = 0.9cm, scale =0.75, xscale=0.7]
\tikzstyle{cpoint}=[shape=semicircle, fill=white,draw=black,minimum width = 0.9cm, scale =0.75, xscale=1, yscale=0.7, shape border rotate = 90,font=\fontsize{14}{16}\selectfont]
\tikzstyle{cocpoint}=[shape=semicircle, fill=white,draw=black,minimum width = 0.9cm, scale =0.75, xscale=1, yscale=0.7, shape border rotate = 270,font=\fontsize{14}{16}\selectfont]

\tikzstyle{every picture}=[baseline=-0.25em,scale=0.5]
\tikzstyle{dotpic}=[] 
\tikzstyle{diredges}=[every to/.style={diredge}]
\tikzstyle{math matrix}=[matrix of math nodes,left delimiter=(,right delimiter=),inner sep=2pt,column sep=1em,row sep=0.5em,nodes={inner sep=0pt},text height=1.5ex, text depth=0.25ex]

\tikzstyle{inline text}=[text height=1.5ex, text depth=0.25ex,yshift=0.5mm]
\tikzstyle{label}=[font=\footnotesize,text height=1.5ex, text depth=0.25ex,yshift=0.5mm]
\tikzstyle{left label}=[label,anchor=east,xshift=1.5mm]
\tikzstyle{right label}=[label,anchor=west,xshift=-1.5mm]

\tikzstyle{braceedge}=[decorate,decoration={brace,amplitude=2mm,raise=-1mm}]
\tikzstyle{small braceedge}=[decorate,decoration={brace,amplitude=1mm,raise=-1mm}]

\tikzstyle{doubled}=[line width=2pt] 
\tikzstyle{boldedge}=[doubled,shorten <=-0.17mm,shorten >=-0.17mm]

\tikzstyle{boldedgedashed}=[very thick,dashed,shorten <=-0.17mm,shorten >=-0.17mm]
\tikzstyle{vboldedgedashed}=[doubled,dashed,shorten <=-0.17mm,shorten >=-0.17mm]
\tikzstyle{left hook arrow}=[left hook-latex]
\tikzstyle{right hook arrow}=[right hook-latex]
\tikzstyle{sembracket}=[line width=0.5pt,shorten <=-0.07mm,shorten >=-0.07mm]

\tikzstyle{causal edge}=[->,thick,gray]
\tikzstyle{causal nondir}=[thick,gray]
\tikzstyle{timeline}=[thick,gray, dashed]

\tikzstyle{cedge}=[<->,thick,gray!70!white]

\tikzstyle{empty diagram}=[draw=gray!40!white,dashed,shape=rectangle,minimum width=1cm,minimum height=1cm]
\tikzstyle{empty diagram small}=[draw=gray!50!white,dashed,shape=rectangle,minimum width=0.6cm,minimum height=0.5cm]

\tikzstyle{dot}=[inner sep=0.7mm,minimum width=0pt,minimum height=0pt,draw,shape=circle]
\tikzstyle{ddot}=[inner sep=0.7mm,doubled, minimum width=2.5mm,minimum height=2.5mm,draw,shape=circle]

\tikzstyle{black dot}=[dot,fill=black]
\tikzstyle{white dot}=[dot,fill=white]
\tikzstyle{green dot}=[white dot] 
\tikzstyle{gray dot}=[dot,fill=gray!40!white]
\tikzstyle{red dot}=[gray dot] 

\tikzstyle{black ddot}=[ddot,fill=black]
\tikzstyle{white ddot}=[ddot,fill=white]
\tikzstyle{gray ddot}=[ddot,fill=gray!40!white]

\tikzstyle{gray edge}=[gray!40!white]

\tikzstyle{small dot}=[inner sep=0.4mm,minimum width=0pt,minimum height=0pt,draw,shape=circle]

\tikzstyle{small black dot}=[small dot,fill=black]
\tikzstyle{small white dot}=[small dot,fill=white]
\tikzstyle{small gray dot}=[small dot,fill=gray!40!white]

\tikzstyle{causal dot}=[inner sep=0.4mm,minimum width=0pt,minimum height=0pt,draw=white,shape=circle,fill=gray!40!white]

\tikzstyle{white phase dot}=[dot,fill=white]
\tikzstyle{white phase ddot}=[ddot,fill=white]

\tikzstyle{gray phase dot}=[dot,fill=gray!40!white]
\tikzstyle{gray phase ddot}=[ddot,fill=gray!40!white]
\tikzstyle{grey phase dot}=[gray phase dot]
\tikzstyle{grey phase ddot}=[gray phase ddot]

\tikzstyle{cnot}=[fill=white,shape=circle,inner sep=-1.4pt]
\tikzstyle{hadamard}=[square box,inner sep=0 pt,font=\tiny\sf,minimum height=3mm,minimum width=3mm]
\tikzstyle{dhadamard}=[hadamard,doubled]
\tikzstyle{antipode}=[white dot,inner sep=0.3mm,font=\footnotesize]

\tikzstyle{scalar}=[diamond,draw,inner sep=0.5pt,font=\small]
\tikzstyle{dscalar}=[diamond,doubled, draw,inner sep=0.5pt,font=\small]

\tikzstyle{small box}=[rectangle,inline text,fill=white,draw,minimum height=5mm,yshift=-0.5mm,minimum width=5mm,font=\small]
\tikzstyle{small gray box}=[small box,fill=gray!30]
\tikzstyle{medium box}=[rectangle,inline text,fill=white,draw,minimum height=5mm,yshift=-0.5mm,minimum width=10mm,font=\small]
\tikzstyle{square box}=[small box] 
\tikzstyle{medium gray box}=[small box,fill=gray!30]
\tikzstyle{large box}=[rectangle,inline text,fill=white,draw,minimum height=5mm,yshift=-0.5mm,minimum width=15mm,font=\small]
\tikzstyle{large gray box}=[small box,fill=gray!30]

\tikzstyle{point}=[regular polygon,regular polygon sides=3,draw,scale=0.75,inner sep=-0.5pt,minimum width=9mm,fill=white,regular polygon rotate=180]
\tikzstyle{copoint}=[regular polygon,regular polygon sides=3,draw,scale=0.75,inner sep=-0.5pt,minimum width=9mm,fill=white]
\tikzstyle{dpoint}=[point,doubled]
\tikzstyle{dcopoint}=[copoint,doubled]

\tikzstyle{tinypoint}=[regular polygon,regular polygon sides=3,draw,scale=0.55,inner sep=-0.15pt,minimum width=6mm,fill=white,regular polygon rotate=180]

\tikzstyle{white point}=[point]
\tikzstyle{green point}=[white point] 
\tikzstyle{white copoint}=[copoint]
\tikzstyle{gray point}=[point,fill=gray!40!white]
\tikzstyle{gray dpoint}=[gray point,doubled]
\tikzstyle{red point}=[gray point] 
\tikzstyle{gray copoint}=[copoint,fill=gray!40!white]
\tikzstyle{gray dcopoint}=[gray copoint,doubled]

\tikzstyle{tiny gray point}=[tinypoint,fill=gray!40!white]

\tikzstyle{diredge}=[->]
\tikzstyle{rdiredge}=[<-]
\tikzstyle{thickdiredge}=[->, very thick]
\tikzstyle{pointer edge}=[->,very thick,gray]
\tikzstyle{pointer edge part}=[very thick,gray]
\tikzstyle{dashed edge}=[dashed]
\tikzstyle{thick dashed edge}=[very thick,dashed]
\tikzstyle{thick gray dashed edge}=[thick dashed edge,gray!90]
\tikzstyle{thick map edge}=[very thick,|->]

\makeatletter
\newcommand{\boxshape}[3]{%
\pgfdeclareshape{#1}{
\inheritsavedanchors[from=rectangle] 
\inheritanchorborder[from=rectangle]
\inheritanchor[from=rectangle]{center}
\inheritanchor[from=rectangle]{north}
\inheritanchor[from=rectangle]{south}
\inheritanchor[from=rectangle]{west}
\inheritanchor[from=rectangle]{east}
\backgroundpath{
\southwest \pgf@xa=\pgf@x \pgf@ya=\pgf@y
\northeast \pgf@xb=\pgf@x \pgf@yb=\pgf@y

\@tempdima=#2
\@tempdimb=#3

\pgfpathmoveto{\pgfpoint{\pgf@xa - 5pt + \@tempdima}{\pgf@ya}}
\pgfpathlineto{\pgfpoint{\pgf@xa - 5pt - \@tempdima}{\pgf@yb}}
\pgfpathlineto{\pgfpoint{\pgf@xb + 5pt + \@tempdimb}{\pgf@yb}}
\pgfpathlineto{\pgfpoint{\pgf@xb + 5pt - \@tempdimb}{\pgf@ya}}
\pgfpathlineto{\pgfpoint{\pgf@xa - 5pt + \@tempdima}{\pgf@ya}}
\pgfpathclose
}
}}

\boxshape{NEbox}{0pt}{5pt}
\boxshape{SEbox}{0pt}{-5pt}
\boxshape{NWbox}{5pt}{0pt}
\boxshape{SWbox}{-5pt}{0pt}
\boxshape{EBox}{-3pt}{3pt}
\boxshape{WBox}{3pt}{-3pt}
\makeatother

\tikzstyle{cloud}=[shape=cloud,draw,minimum width=1.5cm,minimum height=1.5cm]

\tikzstyle{map}=[draw,shape=NEbox,inner sep=2pt,minimum height=6mm,fill=white]
\tikzstyle{mapdag}=[draw,shape=SEbox,inner sep=2pt,minimum height=6mm,fill=white]
\tikzstyle{mapadj}=[draw,shape=SEbox,inner sep=2pt,minimum height=6mm,fill=white]
\tikzstyle{maptrans}=[draw,shape=SWbox,inner sep=2pt,minimum height=6mm,fill=white]
\tikzstyle{mapconj}=[draw,shape=NWbox,inner sep=2pt,minimum height=6mm,fill=white]

\tikzstyle{dbox}=[draw,doubled,shape=rectangle,inner sep=2pt,minimum height=6mm,minimum width=6mm,fill=white]
\tikzstyle{dmap}=[draw,doubled,shape=NEbox,inner sep=2pt,minimum height=6mm,fill=white]
\tikzstyle{dmapdag}=[draw,doubled,shape=SEbox,inner sep=2pt,minimum height=6mm,fill=white]
\tikzstyle{dmapadj}=[draw,doubled,shape=SEbox,inner sep=2pt,minimum height=6mm,fill=white]
\tikzstyle{dmaptrans}=[draw,doubled,shape=SWbox,inner sep=2pt,minimum height=6mm,fill=white]
\tikzstyle{dmapconj}=[draw,doubled,shape=NWbox,inner sep=2pt,minimum height=6mm,fill=white]

\tikzstyle{ddmap}=[draw,doubled,dashed,shape=NEbox,inner sep=2pt,minimum height=6mm,fill=white]
\tikzstyle{ddmapdag}=[draw,doubled,dashed,shape=SEbox,inner sep=2pt,minimum height=6mm,fill=white]
\tikzstyle{ddmapadj}=[draw,doubled,dashed,shape=SEbox,inner sep=2pt,minimum height=6mm,fill=white]
\tikzstyle{ddmaptrans}=[draw,doubled,dashed,shape=SWbox,inner sep=2pt,minimum height=6mm,fill=white]
\tikzstyle{ddmapconj}=[draw,doubled,dashed,shape=NWbox,inner sep=2pt,minimum height=6mm,fill=white]

\boxshape{sNEbox}{0pt}{3pt}
\boxshape{sSEbox}{0pt}{-3pt}
\boxshape{sNWbox}{3pt}{0pt}
\boxshape{sSWbox}{-3pt}{0pt}
\tikzstyle{smap}=[draw,shape=sNEbox,fill=white]
\tikzstyle{smapdag}=[draw,shape=sSEbox,fill=white]
\tikzstyle{smapadj}=[draw,shape=sSEbox,fill=white]
\tikzstyle{smaptrans}=[draw,shape=sSWbox,fill=white]
\tikzstyle{smapconj}=[draw,shape=sNWbox,fill=white]

\tikzstyle{dsmap}=[draw,dashed,shape=sNEbox,fill=white]
\tikzstyle{dsmapdag}=[draw,dashed,shape=sSEbox,fill=white]
\tikzstyle{dsmaptrans}=[draw,dashed,shape=sSWbox,fill=white]
\tikzstyle{dsmapconj}=[draw,dashed,shape=sNWbox,fill=white]

\boxshape{mNEbox}{0pt}{10pt}
\boxshape{mSEbox}{0pt}{-10pt}
\boxshape{mNWbox}{10pt}{0pt}
\boxshape{mSWbox}{-10pt}{0pt}
\tikzstyle{mmap}=[draw,shape=mNEbox]
\tikzstyle{mmapdag}=[draw,shape=mSEbox]
\tikzstyle{mmaptrans}=[draw,shape=mSWbox]
\tikzstyle{mmapconj}=[draw,shape=mNWbox]

\tikzstyle{mmapgray}=[draw,fill=gray!40!white,shape=mNEbox]
\tikzstyle{smapgray}=[draw,fill=gray!40!white,shape=sNEbox]

\makeatletter
\pgfdeclareshape{cornerpoint}{
\inheritsavedanchors[from=rectangle] 
\inheritanchorborder[from=rectangle]
\inheritanchor[from=rectangle]{center}
\inheritanchor[from=rectangle]{north}
\inheritanchor[from=rectangle]{south}
\inheritanchor[from=rectangle]{west}
\inheritanchor[from=rectangle]{east}
\backgroundpath{
\southwest \pgf@xa=\pgf@x \pgf@ya=\pgf@y
\northeast \pgf@xb=\pgf@x \pgf@yb=\pgf@y

\pgfmathsetmacro{\pgf@shorten@left}{\pgfkeysvalueof{/tikz/shorten left}}
\pgfmathsetmacro{\pgf@shorten@right}{\pgfkeysvalueof{/tikz/shorten right}}

\pgfpathmoveto{\pgfpoint{0.5 * (\pgf@xa + \pgf@xb)}{\pgf@ya - 5pt}}
\pgfpathlineto{\pgfpoint{\pgf@xa - 8pt + \pgf@shorten@left}{\pgf@yb - 1.5 * \pgf@shorten@left}}
\pgfpathlineto{\pgfpoint{\pgf@xa - 8pt + \pgf@shorten@left}{\pgf@yb}}
\pgfpathlineto{\pgfpoint{\pgf@xb + 8pt - \pgf@shorten@right}{\pgf@yb}}
\pgfpathlineto{\pgfpoint{\pgf@xb + 8pt - \pgf@shorten@right}{\pgf@yb - 1.5 * \pgf@shorten@right}}
\pgfpathclose
}
}

\pgfdeclareshape{cornercopoint}{
\inheritsavedanchors[from=rectangle] 
\inheritanchorborder[from=rectangle]
\inheritanchor[from=rectangle]{center}
\inheritanchor[from=rectangle]{north}
\inheritanchor[from=rectangle]{south}
\inheritanchor[from=rectangle]{west}
\inheritanchor[from=rectangle]{east}
\backgroundpath{
\southwest \pgf@xa=\pgf@x \pgf@ya=\pgf@y
\northeast \pgf@xb=\pgf@x \pgf@yb=\pgf@y

\pgfmathsetmacro{\pgf@shorten@left}{\pgfkeysvalueof{/tikz/shorten left}}
\pgfmathsetmacro{\pgf@shorten@right}{\pgfkeysvalueof{/tikz/shorten right}}

\pgfpathmoveto{\pgfpoint{0.5 * (\pgf@xa + \pgf@xb)}{\pgf@yb + 5pt}}
\pgfpathlineto{\pgfpoint{\pgf@xa - 8pt + \pgf@shorten@left}{\pgf@ya + 1.5 * \pgf@shorten@left}}
\pgfpathlineto{\pgfpoint{\pgf@xa - 8pt + \pgf@shorten@left}{\pgf@ya}}
\pgfpathlineto{\pgfpoint{\pgf@xb + 8pt - \pgf@shorten@right}{\pgf@ya}}
\pgfpathlineto{\pgfpoint{\pgf@xb + 8pt - \pgf@shorten@right}{\pgf@ya + 1.5 * \pgf@shorten@right}}
\pgfpathclose
}
}

\makeatother

\pgfkeyssetvalue{/tikz/shorten left}{0pt}
\pgfkeyssetvalue{/tikz/shorten right}{0pt}

\tikzstyle{kpoint common}=[draw,fill=white,inner sep=1pt,minimum height=4mm]
\tikzstyle{kpoint}=[shape=cornerpoint,shorten left=5pt,kpoint common]
\tikzstyle{kpoint adjoint}=[shape=cornercopoint,shorten left=5pt,kpoint common]
\tikzstyle{kpoint conjugate}=[shape=cornerpoint,shorten right=5pt,kpoint common]
\tikzstyle{kpoint transpose}=[shape=cornercopoint,shorten right=5pt,kpoint common]
\tikzstyle{kpoint symm}=[shape=cornerpoint,shorten left=5pt,shorten right=5pt,kpoint common]

\tikzstyle{kpointdag}=[kpoint adjoint]
\tikzstyle{kpointadj}=[kpoint adjoint]
\tikzstyle{kpointconj}=[kpoint conjugate]
\tikzstyle{kpointtrans}=[kpoint transpose]

\tikzstyle{dkpoint}=[kpoint,doubled]
\tikzstyle{dkpointdag}=[kpoint adjoint,doubled]
\tikzstyle{dkcopoint}=[kpoint adjoint,doubled]
\tikzstyle{dkpointadj}=[kpoint adjoint,doubled]
\tikzstyle{dkpointconj}=[kpoint conjugate,doubled]
\tikzstyle{dkpointtrans}=[kpoint transpose,doubled]

\tikzstyle{kscalar}=[kpoint common, shape=EBox, inner xsep=-1pt, inner ysep=3pt,font=\small]
\tikzstyle{kscalarconj}=[kpoint common, shape=WBox, inner xsep=-1pt, inner ysep=3pt,font=\small]

 \tikzstyle{upground}=[circuit ee IEC,thick,ground,rotate=90,scale=2.5]
 \tikzstyle{downground}=[circuit ee IEC,thick,ground,rotate=-90,scale=2.5]
 \tikzstyle{bigground}=[regular polygon,regular polygon sides=3,draw=gray,scale=0.50,inner sep=-0.5pt,minimum width=10mm,fill=gray]

\tikzstyle{arrs}=[-latex,font=\small,auto]
\tikzstyle{arrow plain}=[arrs]
\tikzstyle{arrow dashed}=[dashed,arrs]
\tikzstyle{arrow bold}=[very thick,arrs]
\tikzstyle{arrow hide}=[draw=white!0,-]
\tikzstyle{arrow reverse}=[latex-]
\tikzstyle{cdnode}=[]

\tikzstyle{slit}=[line width=2]
\tikzstyle{block}=[line width=4,gray,line cap=round]
\tikzstyle{screen}=[line width=4,black,line cap=round]
\tikzstyle{di}=[diamond,draw,inner sep=0.5pt,font=\small, minimum size = .5cm]
\tikzstyle{sbox}=[rectangle,draw]
\tikzstyle{mirror}=[line width=2,black]
\tikzstyle{trace}=[circuit ee IEC,thick,ground,rotate=0,scale=2]
\tikzstyle{traceState}=[circuit ee IEC,thick,ground,rotate=180,scale=2]
\tikzstyle{detEff}=[circuit ee IEC,thick,ground,rotate=180,scale=1.4]
\tikzstyle{maxMix}=[circuit ee IEC,thick,ground,scale=1.4]
\tikzstyle{particlePath}=[line width=2,gray!40, line cap =round]
\usepackage{fp}
\usetikzlibrary{calc,decorations.pathmorphing,decorations.text,fixedpointarithmetic}

\usepackage{amsmath,amsthm,amsfonts,amssymb}
\usepackage{fontenc}
\usepackage[all]{xy}
\usepackage{graphicx,subfigure,multirow}
\usepackage{tikz}
\usetikzlibrary{shapes, arrows, shadows}
\usepackage[latin1]{inputenc}
\usepackage[scaled]{helvet}
\usepackage[toc, page]{appendix}
\usepackage{xcolor}
\usepackage{comment}

\usepackage{graphicx}				
\usepackage{amssymb}
\newtheorem{definition}{Definition}
\newtheorem{theorem}{Theorem}
\newtheorem{lemma}{Lemma}

\newtheorem*{theorem*}{Theorem}

\begin{document}

\title{Quantum common causes and quantum causal models}  
\author{John-Mark A. Allen}
\affiliation{Department of Computer Science, University of Oxford, Wolfson Building, Parks Road, Oxford OX1 3QD, UK}
\author{Jonathan Barrett}
\affiliation{Department of Computer Science, University of Oxford, Wolfson Building, Parks Road, Oxford OX1 3QD, UK}
\author{Dominic C. Horsman}
\affiliation{Department of Physics, University of Durham, South Road, Durham DH1 3LE, UK}
\author{Ciar\'{a}n M. Lee}
\affiliation{Department of Physics and Astronomy, University College London, Gower Street, London WC1E 6BT, UK}
\author{Robert W. Spekkens}
\affiliation{Perimeter Institute for Theoretical Physics, Waterloo, Ontario N2L 2Y5, Canada}

\begin{abstract}
Reichenbach's principle asserts that if two observed variables are found to be correlated, then there should be a causal explanation of these correlations.  Furthermore, if the explanation is in terms of a common cause, then the conditional probability distribution over the variables given the complete common cause should factorize. The principle is generalized by the formalism of causal models, in which the causal relationships among variables constrain the form of their joint probability distribution.
In the quantum case, however, the observed correlations in Bell experiments cannot be explained in the manner Reichenbach's principle would seem to demand. 
Motivated by this, we introduce a quantum counterpart to the principle.  We demonstrate that under the assumption that quantum dynamics is fundamentally unitary,  if a quantum channel with input $A$ and outputs $B$ and $C$ is compatible with $A$ being a complete common cause of $B$ and $C$, then it must factorize in a particular way.  
Finally, we show how to generalize our quantum version of Reichenbach's principle to a formalism for quantum causal models, and provide examples of how the formalism works. 
\end{abstract}

\maketitle
\tableofcontents

\section{Introduction}

It is a general principle of scientific  thought---and indeed of everyday common sense---that if physical variables are found to be statistically correlated, then there ought to be a causal explanation of this fact. If the dog barks every time the telephone rings, we do not ascribe this to coincidence. A likely explanation is that the sound of the telephone ringing is causing the dog to bark. This is a case where one of the variables is a cause of the other. If sales of ice cream are high on the same days of the year that many people get sunburned, a likely explanation is that the sun was shining on these days and that the hot sun causes both sunburns and the desire to have an ice cream. Here the explanation is not that buying ice cream causes people to get sunburned, nor vice versa, but instead that there is a common cause of both: the hot sun.

That the principle is highly natural is most apparent when it is expressed in its contrapositive form:
if there is {\em no} causal relationship between two variables (i.e. neither is a cause of the other and there is no common cause) then the variables will not be correlated. In particular, without a general commitment to this latter statement, it would be impossible ever to regard two different experiments as independent from one another, or for the results of one scientific team to be regarded as an independent confirmation of the results of another.

This principle of causal explanation was first made explicit by Reichenbach \cite{Reichenbach_book}. 
It is key in scientific investigations which aim to find causal accounts of phenomena from observed statistical correlations.  

Despite the central role of causal explanations in science, there are significant challenges to providing them for the correlations that are observed in quantum experiments~\cite{Wood-Spekkens-15}.  In a Bell experiment, a pair of systems are prepared together, then removed to distant locations where a measurement is implemented on each.  The choice of the measurement made at one wing of the experiment is presumed to be made at space-like separation from that at the other wing. The natural causal explanation of the correlations that one observes in such experiments is that each measurement outcome is influenced by the local measurement setting as well by a common cause located in the joint past of the two measurement events.  But Bell's theorem~\citep{Bell-64} famously rules out this possiblity:
within the standard framework of causal models, if the correlations violate a Bell inequality~\cite{CHSH}---as is predicted by quantum theory and verified experimentally~\cite{hensen2015loophole,shalm2015strong,giustina2015significant}---then a common cause explanation of the correlations is ruled out.   
Furthermore, Ref.~\cite{Wood-Spekkens-15} proves that it is not possible to explain Bell correlations with classical causal models without unwelcome fine-tuning of the parameters.  This includes any attempt to explain Bell correlations with exotic causal influences, such as retrocausality and superluminal signalling.  In the study of classical causation, it is typically assumed that causal explanations should not be fine-tuned \cite{Pearl-09}.

However, the verdict of fine-tuning applies only to {\em classical} models of causation. It was suggested in Ref.~\cite{Wood-Spekkens-15} 
 that it might be possible to provide a satisfactory causal explanation of Bell inequality violations, in particular one that preserves the spirit of Reichenbach's principle and does not require fine-tuning, using a {\em quantum generalisation} of the notion of a causal model. This article seeks to develop such a generalization by first suggesting an intrinsically quantum version of Reichenbach's principle.

Specifically, we consider the case of a quantum system $A$ in the causal past of a bipartite quantum system $BC$ and ask what constraints on the channel from $A$ to $BC$ follows from the assumption that $A$ is the complete common cause of $B$ and $C$. In this scenario we are able to find a natural quantum analogue to Reichenbach's principle. This analogue can be expressed in several equivalent forms, each of which naturally generalises a corresponding classical expression. 
In particular, one of these conditions states that  $A$ is a complete common cause of $BC$ if one can dilate the channel from $A$ to $BC$ to a unitary by introducing two ancillary systems, contained in the causal past of $BC$, such that each ancillary system can influence only one of $B$ and $C$. This unitary dilation codifies the causal relationship between $A$ and $BC$ and illustrates the fact that no other system can influence both $B$ and $C$.
Moreover, our quantum Reichenbach's principle contains the classical version as a special case in the appropriate limit. This suggests that our quantum version is the correct way to generalise Reichenbach's principle.

The mathematical framework of causal models \cite{Pearl-09,Spirtes-Glymour-01} can be seen as a direct generalisation of Reichenbach's principle to arbitrary causal structures. By following this classical example, we are able to generalise our quantum Reichenbach's principle to a framework for quantum causal models. In each case, the original Reichenbach's principle becomes a special case of the framework. Just as with classical causal models, the framework of quantum causal models allows us to analyse the causal structure of arbitrary quantum experiments. It also does so while preserving an appropriate form of Reichenbach's principle (by construction) and avoiding fine-tuning.

Although our main motivation for developing quantum causal models is the possibility of finding a 
satisfactory (i.e., non-fine-tuned) causal explanation of Bell inequality violations~\cite{Wood-Spekkens-15, Chaves-Kueng-15}, they are also likely to have practical applications.  For instance, finding quantum-classical separations in the correlations achievable in novel causal scenarios might lead to new device-independent protocols \cite{Chaves-Majenz-15}, such as randomness extraction and secure key distribution. Quantum causal models may also provide  novel schemes for simulating many body systems in condensed matter physics \cite{Leifer-Poulin-08} and novel means for inferring the underlying causal structure from quantum correlations \cite{fitzsimons2015quantum,Ried-Agnew-15}. 
 
The structure of the paper is as follows. Section~\ref{ccausalnetworks} provides a formal statement of Reichenbach's principle and shows how it can be rigorously justified under certain philosophical assumptions. The main body of results is in Sec.~\ref{qReichenbach}. Here our quantum generalisation of Reichenbach's principle is presented and justified by reasoning parallel to that of the classical case. This is then fleshed out with alternative characterisations of our quantum version of conditional independence and some specific examples. We return to the classical world in Sec.~\ref{ccausalmodels}, discussing classical causal models and providing a rigorous justification of the Markov condition, which plays the role of Reichenbach's principle for general causal structures.
Sec.~\ref{qcausalnetworks} then generalizes these ideas to the quantum sphere, and presents our proposal for quantum causal models. 
 Finally, in Sec.~\ref{priorwork} we describe the relationship of our proposal to prior work on quantum causal models, and in  Sec.~\ref{conclusions} we summarize and describe some directions for future work.

\section{Reichenbach's principle}
\label{ccausalnetworks}

\subsection{Statement}\label{RPstatement}

Reichenbach gave his principle a formal statement 
 in Ref.~\cite{Reichenbach_book}.  Following Ref.~\citep{Cavalcanti-Lal-14},
 we here distinguish two parts of the formalized principle. First is the \emph{qualitative part} which expresses the intuitions described at the beginning of the introduction. The other is the \emph{quantitative part} which constrains the sorts of probability distributions one should assign in the case of a common cause explanation.
 
 The qualitative part of Reichenbach's principle may be stated as follows:
if two physical variables $Y$ and $Z$ are found to be statistically dependent, then there should be a causal explanation of this fact, either: 
\begin{enumerate}
\item $Y$ is a cause of $Z$;
\item $Z$ is a cause of $Y$;
\item there is no causal link between $Y$ and $Z$, but there is a common cause, $X$, influencing $Y$ and $Z$;
\item $Y$ is a cause of $Z$ and there is a common cause, $X$, influencing $Y$ and $Z$; or
\item $Z$ is a cause of $Y$ and there is a common cause, $X$, influencing $Y$ and $Z$.
\end{enumerate}
Note that the causal influences considered here may be indirect (mediated by other variables). 
If none of these causal relations hold between $Y$ and $Z$, then  
we refer to them as {\em ancestrally independent} (because their respective causal ancestries constitute disjoint sets).  Using this terminology, the qualitative part of Reichenbach's principle can be expressed particularly succinctly in its
contrapositive form as: ancestral independence implies statistical independence, i.e., $P(YZ)=P(Y)P(Z)$.  

The quantitative part of Reichenbach's principle applies only to the case where the correlation between $Y$ and $Z$ is due purely to a common cause (case 3 above). 
It states that, in that case, if $X$ is a \emph{complete common cause} for $Y$ and $Z$, meaning that $X$ is the collection of all variables acting as common causes, then $Y$ and $Z$ must be conditionally independent given $X$, so the joint probability distribution $P(XYZ)$ satisfies
\begin{equation}\label{YZCIX}
P(YZ|X) = P(Y|X) P(Z|X).
\end{equation}

\subsection{Justifying the quantitative part of Reichenbach's principle} \label{justifyingclassicalreichenbach}

Within the philosophy of causality, providing an adequate justification of Reichenbach's principle is a delicate issue. It rests on controversy over basic questions, such as what it means for one variable to have a causal influence on another and what is the correct interpretation of probabilistic statements. 
In this section, we discuss one way of justifying the principle, using an assumption of determinism,
which provides a clean motivational story with a natural quantum analogue. Other justifications may be possible.

Suppose we adopt a Bayesian point of view on probabilities: they are the degrees of belief of a rational agent. Dutch book arguments---based on the principle that a rational agent will never accept a set of bets on which they are certain to lose money---can then be given as to why probabilities should be non-negative, sum to $1$ and so forth. But why should an agent who takes $X$ to be a complete common cause for $Y$ and $Z$ arrange their beliefs such that $P(YZ|X) = P(Y|X) P(Z|X)$? 
If the agent does not do this, are they \emph{irrational}?

One way to justify a positive answer to this question is to assume
that in a classical world there is always an underlying deterministic dynamics.   In this case, one variable is causally influenced by another if it has a nontrivial functional dependence upon it in the dynamics. 
Probabilities can be understood as arising merely due to ignorance of the values of unobserved variables. Under these assumptions, one can show that the qualitative part of Reichenbach's principle implies the quantitative part. 

In general, a classical channel describing the influence of random variable $X$ on $Y$ is given by a conditional probability distribution $P(Y|X)$. If we assume underlying deterministic dynamics, then although the value of the variable $Y$ might not be completely determined by the value of $X$, it must be determined by the value of $X$ along with the values of some extra, unobserved, variables in the past of $Y$ which can collectively be denoted $\lambda$. Any variation in the value of $Y$ for a given value of $X$ is then explained by variation in the value $\lambda$. This can be formalised as follows.
\begin{definition}[Classical dilation]\label{defn:classicaldilation}
For a classical channel $P(Y|X)$, a classical deterministic \emph{dilation} is given by some random variable $\lambda$ with probability distribution $P(\lambda)$ and some deterministic function $Y=f(X,\lambda)$  such that
\begin{equation}\label{classicaldilation}
P(Y|X) = \sum_{\lambda} \delta(Y, f(X,\lambda)) P(\lambda),
\end{equation}
where $\delta(X,Y)=1$ if $X=Y$ and $0$ otherwise.  
\end{definition}

\begin{figure}
\includegraphics[scale=0.35,angle=0]{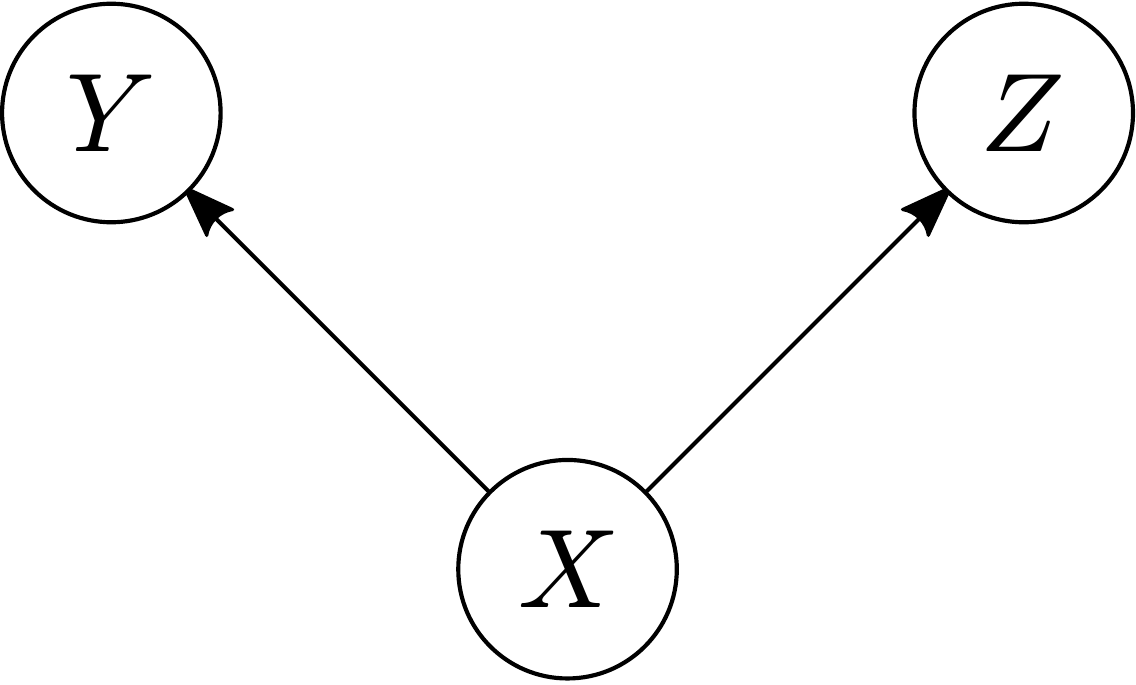}
\caption{A causal structure represented as a directed acyclic graph depicting that $X$ is the complete common cause of $Y$ and $Z$.}\label{xtoyz}
\end{figure}

\begin{figure}
\includegraphics[scale=0.35,angle=0]{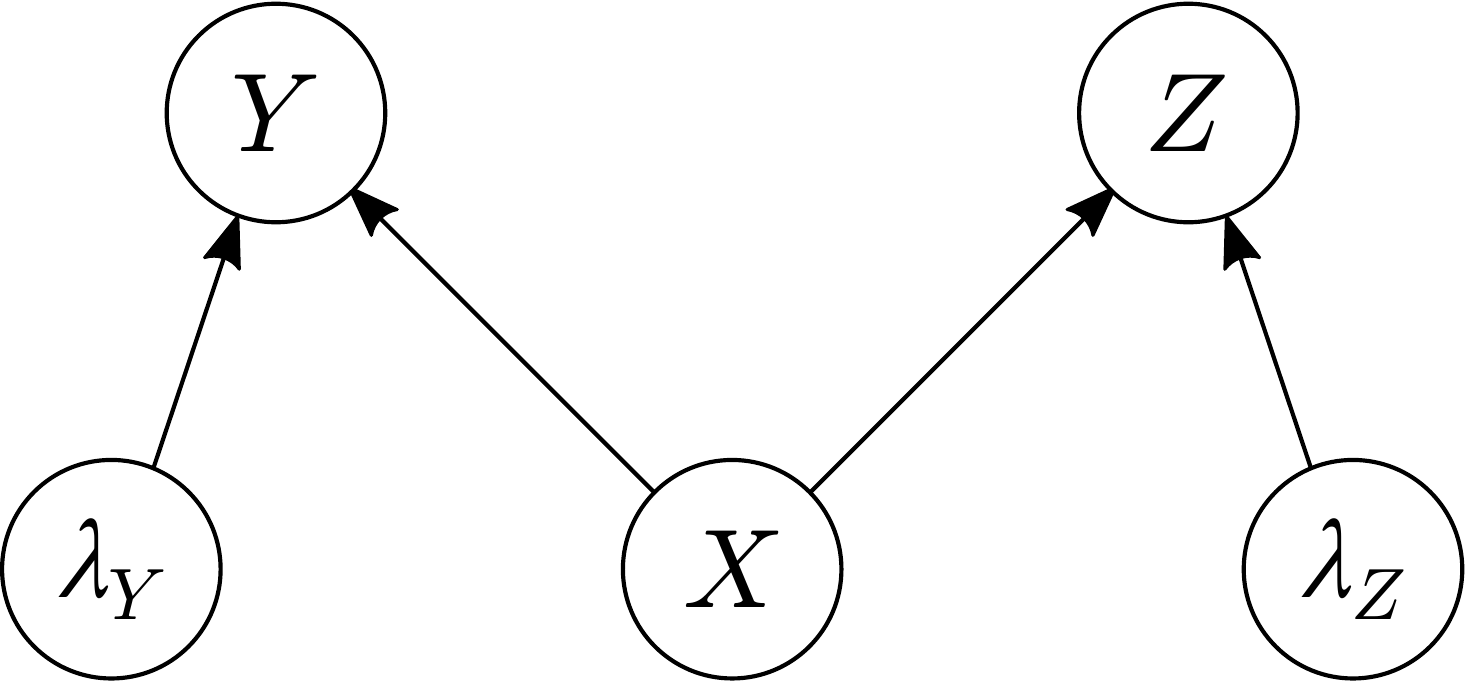}
\caption{The causal structure of Fig.~\ref{xtoyz}, expanded so that $Y$ and $Z$ each has a latent variable as a causal parent in addition to $X$ so that both $Y$ and $Z$ can be made to depend functionally on their parents.}\label{xtoyzdilated}
\end{figure}

We now apply this to the situation depicted in Fig.~\ref{xtoyz}, where $X$ is the complete common cause of $Y$ and $Z$.  The conditional distribution $P(YZ|X)$ admits of a dilation in terms of an ancillary unobserved variable $\lambda$, for some distribution $P(\lambda)$ and a function $f = (f'_Y,f'_Z)$ from $(\lambda,X)$ to $(Y,Z)$ such that $Y = f'_Y(\lambda,X)$ and $Z = f'_Z(\lambda,X)$.  The assumption that $X$ is the \emph{complete} common cause of $Y$ and $Z$ implies that the ancillary variable $\lambda$ can be split into a pair of ancestrally independent variables, $\lambda_Y$ and $\lambda_Z$, where $\lambda_Y$ only influences $Y$ and $\lambda_Z$ only influences $Z$ \footnote{This is because any other $\lambda$ would necessarily introduce new common causes for $Y$ and $Z$ that are not screened through $X$, which would violate the assumption that $X$ is a complete common cause.}.
  It follows that there must exist $\lambda_Y$ and $\lambda_Z$ that are causally related to $X$, $Y$ and $Z$ as depicted in Fig.~\ref{xtoyzdilated}, where the causal dependences are deterministic and given by a pair of functions $f_Y$ and $f_Z$ such that $Y = f_Y(\lambda_Y,X)$ and $Z = f_Z(\lambda_Z,X)$.

In this case, we have
\begin{align}\label{pyzx}
&P(YZ|X)\nonumber\\
&= \sum_{\lambda_Y,\lambda_Z} \delta(Y,f_Y(\lambda_Y,X)) \delta(Z,f_Z(\lambda_Z,X)) P(\lambda_Y,\lambda_Z)
\end{align}
   Finally, given the qualitative part of Reichenbach's principle, the ancestral independence of $\lambda_Y$ and $\lambda_Z$ in the causal structure implies that $P(\lambda_Y,\lambda_Z)= P(\lambda_Y)P(\lambda_Z)$.  It then follows that $P(YZ|X) = P(Y|X) P(Z|X)$, which establishes the quantitative part of Reichenbach's principle.

A well-known converse statement is also worth noting:
any classical channel $P(YZ|X)$ satisfying $P(YZ|X)=P(Y|X)P(Z|X)$ admits of a dilation where $X$ is the complete common cause of $Y$ and $Z$ \cite{Pearl-09}.  

Summarizing, we can identify what it means for $P(YZ|X)$ to be explainable in terms of $X$ being a complete common cause of $Y$ and $Z$ by appealing to the qualitative part of Reichenbach's principle and fundamental determinism.  The definition can be formalized into a mathematical condition as follows:

\begin{definition}[Classical compatibility]\label{classicalcompatibility}
$P(YZ|X)$ is said to be {\em compatible} with $X$ being the complete common cause of $Y$ and $Z$ if one can find variables  $\lambda_Y$ and $\lambda_Z$, distributions $P(\lambda_Y)$ and $P(\lambda_Z)$, a function $f_Y$ from $(\lambda_Y, X)$ to $Y$ and a function $f_Z$ from $(\lambda_Z, X)$ to $Z$, 
such that these constitute a dilation of $P(YZ|X)$, that is, such that
\begin{align}
&P(YZ|X)\nonumber\\
&= \sum_{\lambda_Y,\lambda_Z} \delta(Y,f_Y(\lambda_Y,X)) \delta(Z,f_Z(\lambda_Z,X)) P(\lambda_Y) P(\lambda_Z)
\end{align}
\end{definition}
\color{black}

With this definition, we can summarize the result described above as follows.

\begin{theorem}\label{Cversionmaintheorem}
Given a conditional probability distribution $P(YZ|X)$, the following are equivalent:
\begin{enumerate}
\item $P(YZ|X)$ is compatible with $X$ being the complete common cause of $Y$ and $Z$.
\item $P(YZ|X) = P(Y|Z)P(Z|X)$.
\end{enumerate}
\end{theorem}

The $1 \to 2$ implication is what establishes that a rational agent should espouse the quantitative part of Reichenbach's principle if they espouse the qualitative part and fundamental determinism.

The $2 \to 1$ implication allows one to deduce a possible causal explanation of an observed distribution from a feature of that distribution. However, it is important to stress that it only establishes a \emph{possible} causal explanation. It does not state that this is the \emph{only} causal explanation.  Indeed, it may be possible to satisfy this conditional independence relation within alternative causal structures by fine-tuning the strengths of the causal dependences. 
 However, as noted above, fine-tuned causal explanations are typically rejected as bad explanations in the field of causal inference. Therefore, the best explanation of the conditional independence of $Y$ and $Z$ given $X$ is that $X$ is the complete common cause of $Y$ and $Z$.

\section{The quantum version of Reichenbach's principle}\label{qReichenbach}

In this section, we introduce our quantum version of Reichenbach's principle.  The definition of a quantum causal model that we provide in Sec.~\ref{qcausalnetworks} can be seen as generalizing these ideas in much the same way that classical causal models generalize the classical version of Reichenbach's principle.

\subsection{Quantum preliminaries}
For simplicity, we assume throughout that all quantum systems are finite-dimensional. Given a quantum system $A$, we will write $\mathcal{H}_A$ for the corresponding Hilbert space, $d_A$ for the dimension of $\mathcal{H}_A$, and $I_A$ for the identity on $\mathcal{H}_A$. We will also write $\mathcal{H}_A^\ast$ for the dual space to $\mathcal{H}_A$, and $I_{A^\ast}$ for the identity on the dual space. If a quantum system is initially uncorrelated with any other system, then the most general time evolution of the system corresponds to a quantum channel, i.e., a completely postive trace-preserving (CPTP) map. If the system at the initial time is labelled $A$, with Hilbert space $\mathcal{H}_A$, and the system at the later time is labelled $B$, with Hilbert space $\mathcal{H}_B$, then the CPTP map is
\begin{equation}
\mathcal{E}_{B|A}: \mathcal{L}(\mathcal{H}_A) \rightarrow \mathcal{L}(\mathcal{H}_B),
\end{equation}
where $\mathcal{L}(\mathcal{H})$ is the set of linear operators on $\mathcal{H}$. 

An alternative way to express the channel $\mathcal{E}_{B|A}$ is as an operator, using a variant of the Choi-Jamio\l{}kowski isomorphism \cite{Jamiolkowski1972,Choi1975}:
\begin{equation} \label{ChoiJam}
\rho_{B|A} := \sum_{ij} \mathcal{E}_{B|A} (|i\rangle_A\langle j| )\otimes  |i\rangle_{A^\ast}\langle j| .
\end{equation}
Here, the vectors $\{ |i\rangle_A \}$ form an orthonormal basis of the Hilbert space $\mathcal{H}_A$. 
The vectors $\{ |i\rangle_{A^\ast} \}$ form the dual basis, belonging to $\mathcal{H}_A^{\ast}$. The operator $\rho_{B|A}$ therefore acts on the Hilbert space $\mathcal{H}_B \otimes \mathcal{H}_A^\ast$. Although the expression above involves an arbitrary choice of orthonormal basis, the operator $\rho_{B|A}$ thus defined is independent of the choice of basis. This version of the Choi-Jamio\l{}kowski isomorphism was chosen because it is both basis-independent and a positive operator.
Following \cite{Leifer-Spekkens-13}, we have chosen the operator $\rho_{B|A}$ to be normalized in such a way that
 $\mathrm{Tr}_B(\rho_{B|A}) = I_{A^\ast}$ (in analogy with the normalization condition $\sum_Y P(Y|X) = 1$ for a classical channel $P(Y|X)$).

Suppose that $\rho_B = \mathcal{E}_{B|A}(\rho_A)$.  Given that the operator $\rho_{B|A}$ contains all of the information about the channel $\mathcal{E}_{B|A}$, the question arises of how one can express $\rho_B$ in terms of $\rho_{B|A}$ and $\rho_A$. 
Recall that $\rho_{B|A}$ is defined on $\mathcal{H}_{B}\otimes \mathcal{H}_{A^*}$, while $\rho_A$ is defined on $\mathcal{H}_{A}$.  As we discuss further in Sec.~\ref{qcausalnetworks}, by defining an appropriate ``linking operator'' on $\mathcal{H}_{\bf A} := \mathcal{H}_{A}\otimes \mathcal{H}_{A^*}$,
\begin{equation}\label{linkingoperator}
\tau^{\rm id}_{\bf A} := \sum_{lm} |l\rangle_{A^*} \langle m|\otimes|l\rangle_A \langle m|
\end{equation}
where $\{  | l \rangle_{A}\}_l $ and $\{  | l \rangle_{A^*}\}_l $ are orthonormal bases on $\mathcal{H}_{A}$ and $\mathcal{H}_{A^*}$ respectively, 
one can write $\rho_B  = Tr_{\bf A} ( \rho_{B|A} \tau^{\rm id}_{\bf A} \rho_A  )$.  This expression is meant to be reminiscent of the classical formula $P(Y) =\sum_Y P(Y|X) P(X)$.

Given an operator $\rho_{AB\cdots|CD\cdots}$, acting on $\mathcal{H}_A\otimes \mathcal{H}_B\otimes\cdots \otimes \mathcal{H}_{C^\ast}\otimes \mathcal{H}_{D^\ast}\otimes\cdots  $, we will use the same expression  with missing indices  to denote the result of taking partial traces on the corresponding factor spaces. For example, given a channel $\rho_{AB|CD}$, we write $\rho_{A|CD} := \mathrm{Tr}_{B}(\rho_{AB|CD})$. 

When writing products of operators, we will sometimes suppress tensor products with identities. For example, 
$(\rho_{B|A} \otimes I_C) (\rho_{C|A} \otimes I_B)$ will be written simply as  $\rho_{B|A} \rho_{C|A}$.

\subsection{Main result}\label{qcommoncauses}

The qualitative part of Reichenbach's principle can be applied to quantum theory with almost no change: if quantum systems $B$ and $C$ are correlated then this must have a causal explanation in one of the five forms listed in Sec.~\ref{RPstatement} (except with classical variables $X$, $Y$ and $Z$ replaced by quantum systems $A$, $B$ and $C$). Here,  for two quantum systems to be correlated means that their joint quantum state does not factorize.

\begin{figure}
\includegraphics[scale=0.35,angle=0]{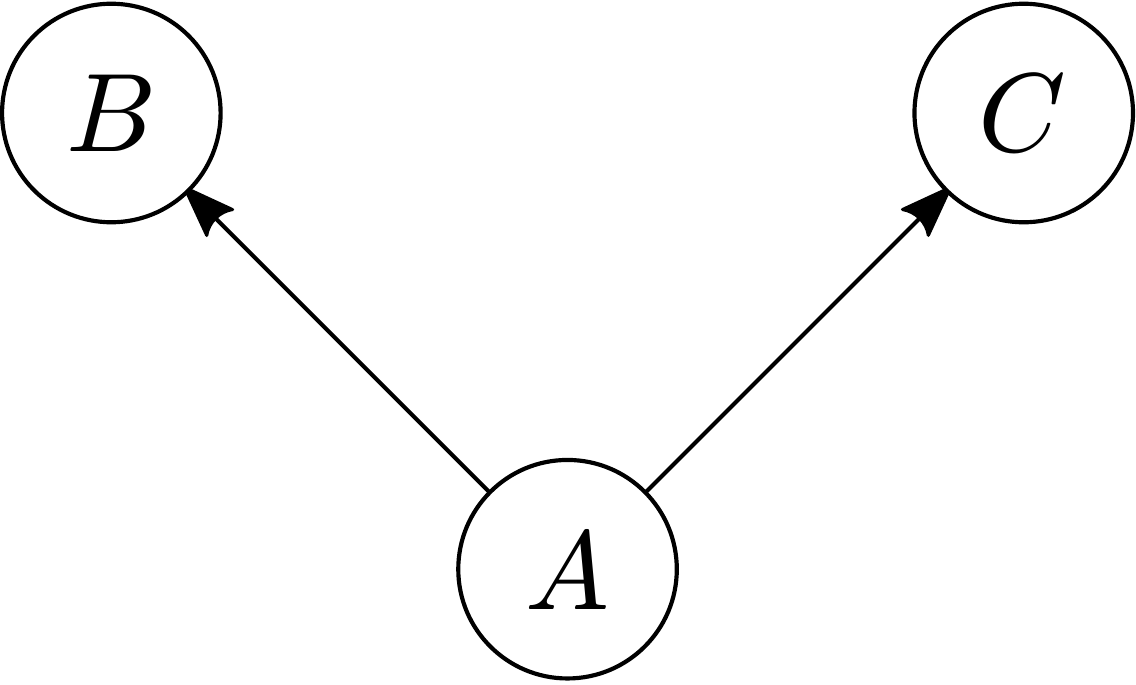}
\caption{A causal structure relating three quantum systems with $A$ the complete common cause of $B$ and $C$.}\label{atobc}
\end{figure}

Finding a quantum version of the {\em quantitative} part of Reichenbach's principle is more subtle. 
If a quantum system $A$ is a complete common cause of $B$ and $C$ (as depicted in Fig.~\ref{atobc}), then one expects there to be some constraint analogous to the classical constraint that $P(YZ|X) = P(Y|X)P(Z|X)$.  If one tries to do this by generalising the joint distribution $P(XYZ)$, then one immediately faces the problem that textbook quantum theory has no analogue of a joint distribution for a collection of quantum systems in which some are causal descendants of others.   The situation is improved if one focusses on finding an analogue of $P(YZ|X)$ instead.  In standard quantum theory, as long as a system $A$ is initially uncorrelated with its environment, then the evolution from $A$ to $BC$ is described by a channel $\mathcal{E}_{BC|A}$.  The operator that is isomorphic to this channel by Eq.~\eqref{ChoiJam}, denoted $\rho_{BC|A}$, seems to be a natural analogue of $P(YZ|X)$.  However, even in this case, it is not obvious what constraint on $\rho_{BC|A}$ should serve as the analogue of the classical constraint  $P(YZ|X) = P(Y|X)P(Z|X)$.  

The treatment of generic causal networks of quantum systems is deferred to the full definition of quantum causal models in Sec.~\ref{qcausalnetworks}. This section focuses on the case of a channel $\rho_{BC|A}$.

In Sec.~\ref{justifyingclassicalreichenbach}, we demonstrated how to justify the quantitative part of Reichenbach's principle from the qualitative part in the classical case under the assumption that all dynamics are fundamentally deterministic.  We shall now make an analogous argument in the quantum case by assuming that quantum dynamics are fundamentally \emph{unitary}. Just as in the classical case, this assumption simply provides a clean way to motivate our result and alternative justifications may be possible.

In general, a quantum channel from $A$ to $B$ is given by a CPTP map $\mathcal{E}_{B|A}$. Assuming underlying unitary dynamics, then the output state at $B$ must depend unitarily on $A$ along with some extra ancillary system $\lambda$ in the past of $B$. This can be formalised as follows.

\begin{definition}[Unitary dilation]
For a quantum channel $\mathcal{E}_{B|A}$ a quantum \emph{unitary dilation} is given by some ancillary quantum system $\lambda$ with state $\rho_{\lambda}$ and some unitary $U$ from  $\mathcal{H}_{A}\otimes \mathcal{H}_{\lambda}$ to $\mathcal{H}_{B}\otimes \mathcal{H}_{\bar{B}}$ such that
\[
\mathcal{E}_{B|A} (\cdot )= \mathrm{Tr}_{\bar{B}} \left( U (\cdot \otimes \rho_{\lambda}) U^{\dag} \right),
\]
where the dimension of $\bar{B}$ is fixed by the requirement for unitarity that $d_A d_\lambda = d_B d_{\bar{B}}$.
\end{definition}

If we represent the channels by our variant of the Choi-Jamio{\l}kowski isomorphism, Eq.~\eqref{ChoiJam}, with $\rho_{B|A}$ representing $\mathcal{E}_{B|A}$ and $\rho^{U}_{B\bar{B}|A \lambda}$ representing $U(\cdot) U^{\dag}$,
 then the dilation equation has the form
\[
\rho_{B|A} = \mathrm{Tr}_{\bar{B} \boldsymbol{\lambda}} \left( \rho^{U}_{B\bar{B}|A \lambda} \tau^{\rm id}_{\boldsymbol{\lambda}}  \rho_{\lambda} \right)
\]
where $\tau^{\rm id}_{\boldsymbol{\lambda}}$ is the linking operator defined in Eq.~\eqref{linkingoperator}.

Just as in the classical case, we would like to apply this to the situation depicted in Fig.~\ref{atobc}, where $A$ is the complete common cause for $B$ and $C$. This was easy classically as it is clear what it means for a classical variable, $X$, to have no causal influence on another, $Y$, in a deterministic system. Specifically, if the collection of inputs other than $X$ is denoted  $\bar{X}$ so there is a deterministic function $f$ such that $Y=f(X,\bar{X})$, then the assumption that $X$ has no causal influence on $Y$ is formalized as $f(X,\bar{X})=f'(\bar{X})$ for some function $f'$. In unitary quantum theory the corresponding condition is less obvious, so we spell it out explicitly with a definition.

\begin{definition}[No influence]\label{nonsig}
Consider a unitary channel $\rho^{U}_{B\bar{B}|A\bar{A}}$ from $A\bar{A}$ to $B\bar{B}$. $A$ has \emph{no causal influence} on $B$ if and only if for $\rho_{B | A\bar{A}} := {\rm tr}_{\bar{B}} \rho^{U}_{B\bar{B}|A\bar{A}}$, we have
 $\rho_{B | A\bar{A}} = I_{A^\ast}\otimes \rho_{B|\bar{A}}$.
\end{definition}

An equivalent definition is this: $A$ has no causal influence on $B$ in some unitary channel if and only if the marginal output state at $B$ is independent of any operations performed on $A$ before the $A$ system enters the channel. There is a rich literature concerning similar properties of unitary operators from various perspectives. In particular, the results of Ref.~\cite{Schumacher-Westmoreland-05} are very close to ours (where they use the phrase ``nonsignalling'' rather than ``no causal influence'') and Refs.~\cite{Beckman-Gottesman-01,eggeling2002semicausal} contain similar results (where they say ``semi-causal'' rather than ``no causal influence'').

We can now apply this to the complete common cause situation of Fig.~\ref{atobc}. The channel $\mathcal{E}_{BC|A}$ admits a unitary dilation in terms of an ancillary system $\lambda$, for some state $\rho_\lambda$ and unitary $U$ from $\lambda A$ to $BDC$. Here, an ancillary output $D$ is generally required so that dimensions of inputs and outputs match, but is not important and will always be traced out. This dilation is such that $\mathcal{E}_{BC|A} (\cdot )= \mathrm{Tr}_{D} \left( U (\cdot \otimes \rho_{\lambda}) U^{\dag} \right)$. 

\begin{figure}
\includegraphics[scale=0.35,angle=0]{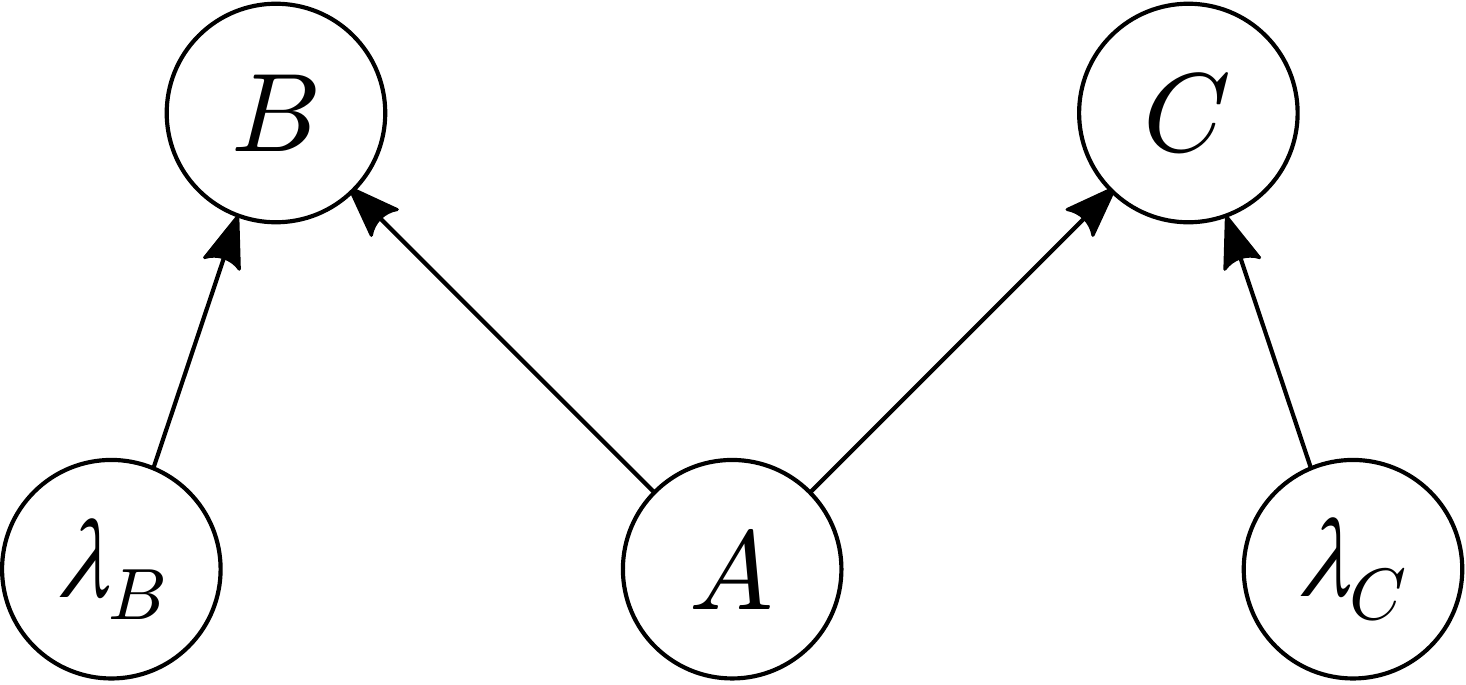}
\caption{The causal structure of Fig.~\ref{atobc}, expanded so that $B$ and $C$ each has a latent system as a causal parent in addition to $A$. By analogy the classical case, we take $B$ and $C$ to depend unitarily on their $\lambda_B$, $A$, and $\lambda_C$.}\label{unitary}
\end{figure}

Just as in Sec.~\ref{justifyingclassicalreichenbach}, the assumption that $A$ is a \emph{complete} common cause for $B$ and $C$ implies that the ancilla $\lambda$ can be factorized into ancestrally independent $\lambda_B$ and $\lambda_C$ where $\lambda_B$ has no causal influence on $C$ and $\lambda_C$ has no causal influence on $B$. It follows that systems $\lambda_B$ and $\lambda_C$ are causally related to $A$, $B$, and $C$ as depicted in Fig.~\ref{unitary}.

The ancestral independence of $\lambda_B$ and $\lambda_C$ implies that the quantum state on $\lambda$ factorizes across the $\lambda_B$, $\lambda_C$ partition, $\rho_{\lambda}=\rho_{\lambda_B}\rho_{\lambda_C}$,
suggesting the following quantum analogue to our classical compatibility condition of Def.~\ref{classicalcompatibility}.

\begin{definition}[Quantum compatibility]\label{Qcompatibility} $\rho_{BC|A}$ is said to be compatible with $A$ being a complete common cause of $B$ and $C$, if it is possible to find ancillary quantum systems $\lambda_B$ and $\lambda_C$, states $\rho_{\lambda_B}$ and $\rho_{\lambda_C}$, and a unitary channel 
where $\lambda_B$ has no causal influence on $C$ and $\lambda_C$ has no causal influence on $B$, such that these constitute a dilation of $\rho_{B C|A}$.
\end{definition}

All that remains is to show that this, together with the qualitative part of the quantum Reichenbach's principle, implies an appropriate quantitative part (generalising Thm~\ref{Cversionmaintheorem}).

\begin{theorem}\label{maintheorem}
The following are equivalent:
\begin{enumerate}
\item 
$\rho_{BC|A}$ is compatible with $A$ being the complete common cause of $B$ and $C$.
\item $\rho_{BC|A} = \rho_{B|A} \rho_{C|A}$.
\end{enumerate}
\end{theorem}

The proof is in Appendix~\ref{maintheoremproof}. Note that there is no ordering ambiguity on the right-hand side of the second condition, because the two terms must commute.  This is seen by taking 
 the Hermitian conjugate of both sides of the equation and recalling that $\rho_{BC|A}$ is Hermitian.

The strong analogy that exists between Thms~\ref{Cversionmaintheorem} and \ref{maintheorem} suggests the following definition:
\begin{definition}[Quantum conditional independence of outputs given input]\label{QCIoutgivenin}
Given a quantum channel $\rho_{BC|A}$, the outputs are said to be \emph{quantum conditionally independent} given the input if and only if $\rho_{BC|A} = \rho_{B|A} \rho_{C|A}$.
\end{definition}
\color{black}
It is easily seen that the quantum definition reduces to the classical definition in the case that the channel $\rho_{BC|A}$ is invariant under the operation of completely dephasing the systems $A$, $B$, and $C$ in some basis.  More precisely: if fixed bases are chosen for $\mathcal{H}_A, \mathcal{H}_B, \mathcal{H}_C$, and the operator $\rho_{BC|A}$ is diagonal when written with respect to the tensor product of these bases, then the outputs are quantum conditionally independent given the input if and only if the classical channel defined by the diagonal elements of the matrix has the property that the outputs are conditionally independent given the input.

With this terminological convention in hand, we can express our quantum version of the quantitative part of Reichenbach's principle as follows: if a channel $\rho_{BC|A}$ is compatible with $A$ being a complete common cause of $B$ and $C$, then for this channel, $B$ and $C$ are quantum conditionally independent given $A$.  

The $1 \to 2$ implication in the theorem is what establishes 
the quantum version of the quantitative part of Reichenbach's principle.

The $2 \to 1$ implication is pertinent to causal inference: analogously to the classical case, if one grants the implausibility of fine-tuning, then one must grant that the {\em most plausible} explanation of the quantum conditional independence of outputs $B$ and $C$ given input $A$ is that $A$ is a complete common cause of $B$ and $C$. 
 \color{black}

Thm~\ref{maintheorem}, and the surrounding discussion, motivates the definition of quantum causal models given in Sec.~\ref{qcausalnetworks}. For the rest of this section we make some further remarks about the quantum version of Reichenbach's principle. 

\subsection{Alternative expressions for quantum conditional independence of outputs given input}\label{AltExpCI}

Classically, conditional independence of $Y$ and $Z$ given $X$ is standardly expressed as $P(YZ|X)=P(Y|X)P(Z|X)$.  However, there are alternative ways of expressing this constraint.  

For instance, if one defines the joint distribution over $X, Y, Z$ that one obtains by feeding the uniform distribution on $X$ into the channel $P(YZ|X)$---that is, $\hat{P}(XYZ) := P(YZ|X)\frac{1}{d_X}$, where $d_X$ is the cardinality of $X$---then $Y$ and $Z$ being conditionally independent given $X$ in $P(YZ|X)$ can be expressed as the vanishing of the conditional mutual information of $Y$ and $Z$ given $X$ in the distribution $\hat{P}(XYZ)$ \cite{Pearl-09}.  This conditional mutual information is defined as $I(Y:Z|X) := H(Y,X) + H(Z,X) - H(X,Y,Z) - H(X)$, with $H(\cdot)$ denoting the Shannon entropy of the marginal on the subset of variables indicated in its argument.  Therefore, 
the condition is simply $I(Y:Z|X)=0$.

Similarly, if $Y$ and $Z$ are conditionally independent given $X$ in $P(YZ|X)$, then it is possible to mathematically represent the channel $P(YZ|X)$ as the following sequence of operations: copy $X$, then process one copy into $Y$ via the channel $P(Y|X)$ and process the other into $Z$ via the channel $P(Z|X)$.   

We present here the quantum analogues of these alternative expressions.  They will be found to be useful for developing intuitions about quantum conditional independence and in proving Thm~\ref{maintheorem}.
Recall that the quantum conditional mutual information of $B$ and $C$ given $A$ is defined as $I(B:C|A) := S(B,A) + S(C,A) - S(A,B,C) - S(A)$, where $S(\cdot)$ denotes the von Neumann entropy of the reduced state on the subsystem that is specified by its argument.  Analogously to the classical case, we will use a hat to denote an operator renormalized such that the trace is $1$. For example, if $\rho_{B|A}$ is the operator representing a channel from $A$ to $B$, then $\hat{\rho}_{B|A} := (1/d_A)\rho_{B|A}$.

\begin{theorem}\label{alternativeexpressions}
Given a channel $\rho_{BC|A}$, the following conditions are also equivalent to the quantum conditional independence of the outputs given the input (condition 2 of Thm~\ref{maintheorem}):
\begin{enumerate}
\item[3.] $I(B:C|A) = 0$ where  $I(B:C|A)$ is the quantum conditional mutual information of $B$ and $C$ given $A$ evaluated on the (positive, trace-one) operator $\hat{\rho}_{BC|A}:= (1/d_A)\rho_{BC|A}$.
\item[4.] The Hilbert space for the $A$ system can be decomposed as $\mathcal{H}_{A} = \bigoplus_i \mathcal{H}_{A_i^L}\otimes\mathcal{H}_{A_i^R}$ and $\rho_{BC|A} = \sum_i \left(\rho_{B|A_i^L}\otimes\rho_{C|A_i^R}\right)$, where for each $i$, $\rho_{B|A_i^L}$ represents a CPTP map $\mathcal{B}(\mathcal{H}_{A_i^L}) \rightarrow \mathcal{B}(\mathcal{H}_B)$, and $\rho_{C|A_i^R}$ a CPTP map $\mathcal{B}(\mathcal{H}_{A_i^R}) \rightarrow \mathcal{B}(\mathcal{H}_C)$.
\end{enumerate}
\end{theorem}
The proof is in Appendix~\ref{maintheoremproof}. That conditions 3 and 4 are equivalent follows as a corollary of Thm~6 of Ref.~\cite{Hayden-Jozsa-04}.  Our main contribution is showing that these are also equivalent to condition 2 of Thm~\ref{maintheorem}.

The final condition can be described as follows. First one imagines decomposing the system $A$ into a direct sum of subspaces, each of which is denoted $A_i$. For each $i$, the subspace $A_i$ is split into two factors, denoted  $A_i^L$ and $A_i^R$, with one factor evolving via a channel $\rho_{B|A_i^L}$ into system $B$, and the other factor evolving via $\rho_{C|A_i^R}$ into system $C$. 
In the special case where there is only a single value of $i$, this is simply a factorization of the $A$ system into two parts.  In the special case where all of the  $A_i^L$ and $A_i^R$ are 1-dimensional Hilbert spaces, it is simply an incoherent copy operation.

\subsection{Circuit representations}\label{remarks}

It is instructive to summarize the contents of Thms~\ref{Cversionmaintheorem} and \ref{maintheorem} using circuit diagrams.   

\begin{figure}[h!]
    \includegraphics[scale=0.35,angle=0]{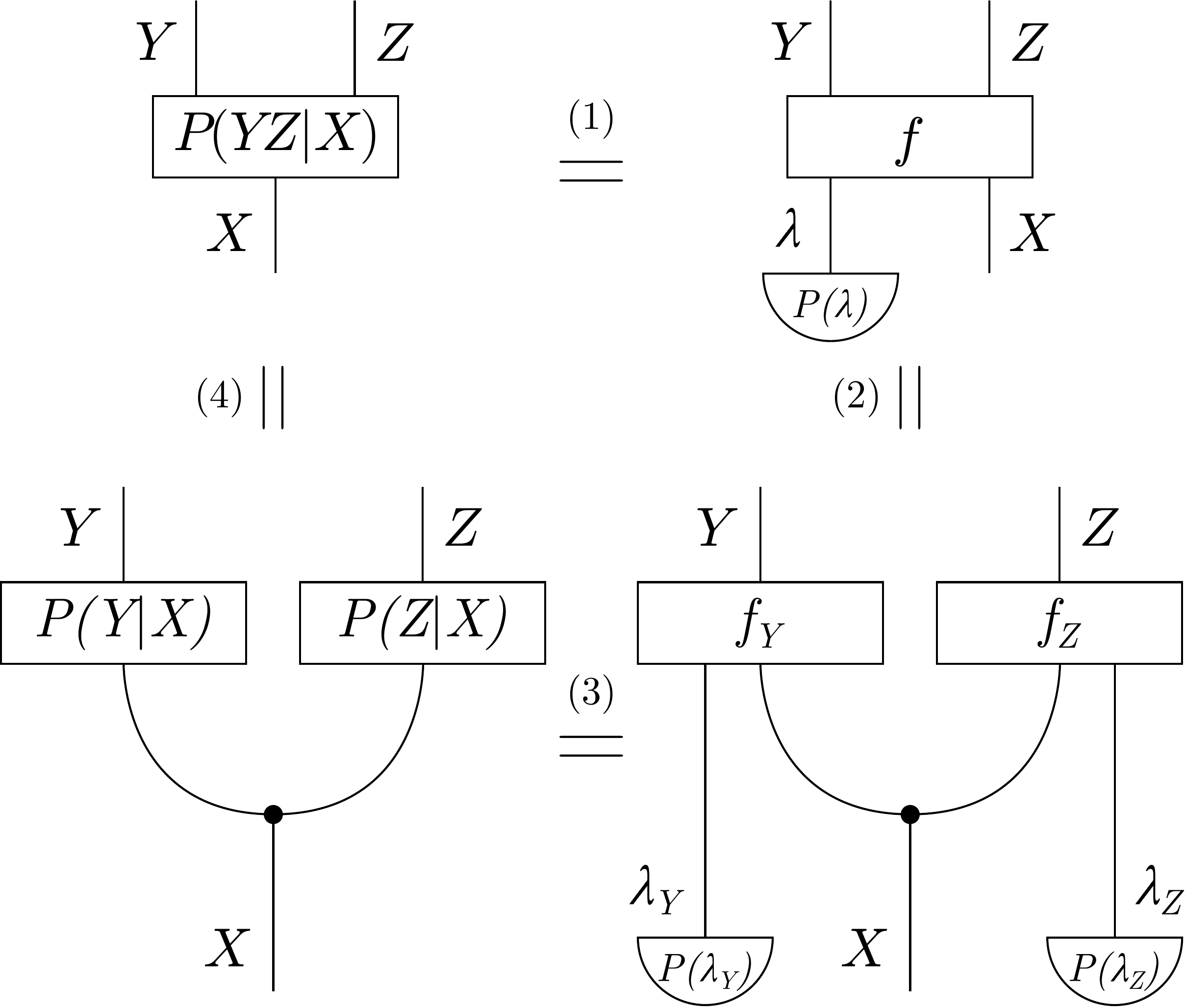}
    \caption{\label{CCircuits} Diagrammatic representation of Thm~\ref{Cversionmaintheorem} and of alternative expressions for conditional independence of outputs given input (the classical analogue of Thm~\ref{alternativeexpressions}).}
\end{figure}

The classical case is shown in Fig.~\ref{CCircuits}, where four equivalent circuits represent the action of a channel $P(YZ|X)$, for which the outputs $YZ$ are conditionally independent given the input $X$. The dot in the lower two circuits represents a classical copy operation. Equality (1) simply asserts that the conditional probability distribution $P(YZ|X)$ admits a classical dilation, as in Def.~\ref{defn:classicaldilation}.  Equality (4) asserts that the channel is equivalent to a sequence of operations in which $X$ is copied, with one copy the input to a channel $P(Y|X)$ and one copy the input to a channel $P(Z|X)$. As discussed at the beginning of Sec.~\ref{AltExpCI}, this is one way of expressing the fact that $Y$ and $Z$ are conditionally independent given $X$. Equality (3) asserts that $P(Y|X)$ and $P(Z|X)$ separately admit classical dilations. Finally, equality (2) asserts that $P(YZ|X)$ is compatible with $X$ being a complete common cause of $Y$ and $Z$ by depicting conditions under which $\lambda_Y$ has no influence on $Z$ and $\lambda_Z$ has no influence on $Y$.
\begin{figure}[h!]
    \includegraphics[scale=0.35,angle=0]{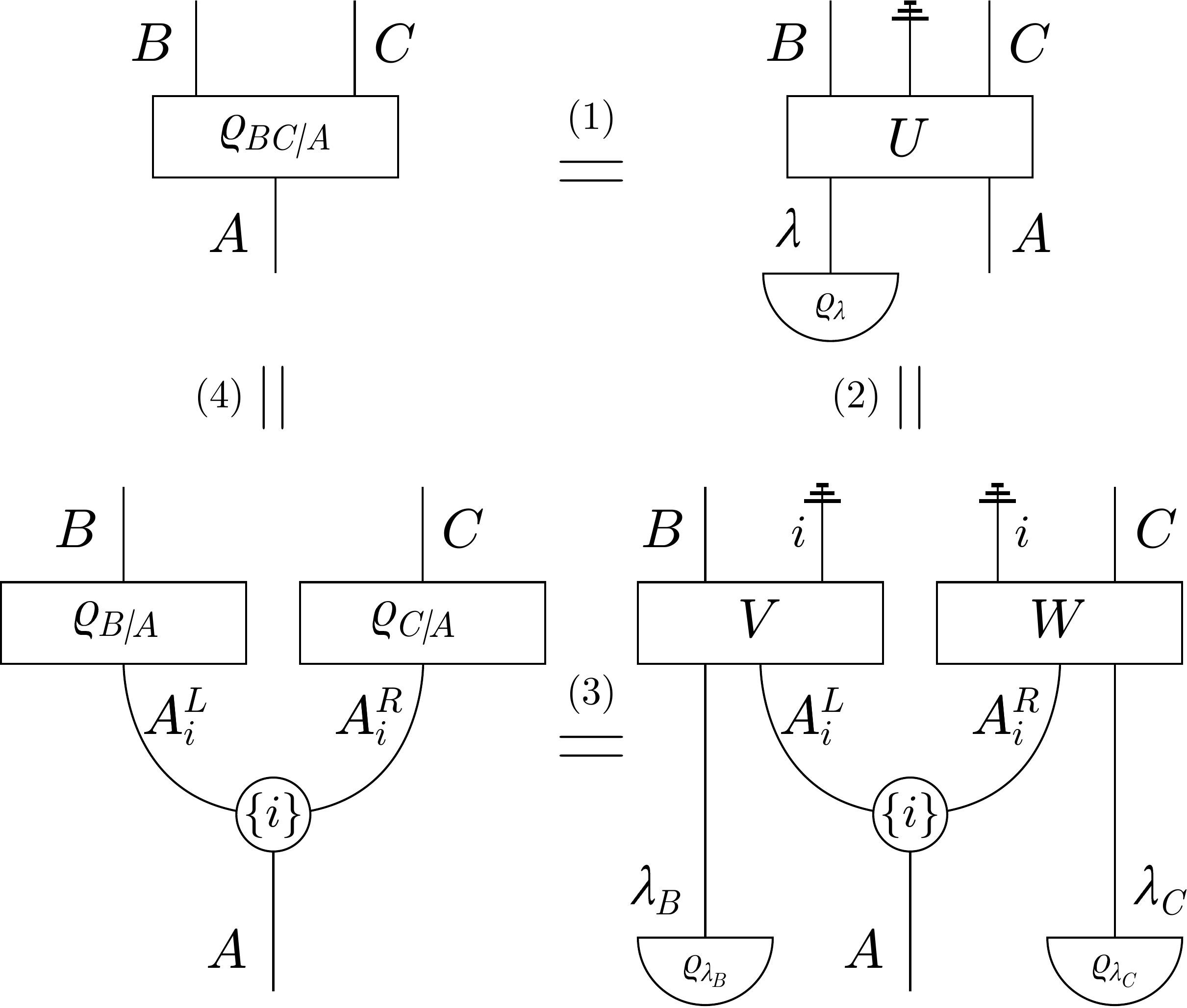}
    \caption{\label{QCircuits} Diagrammatic representation of Thm~\ref{maintheorem} and of alternative expressions for quantum conditional independence of outputs given input (Thm~\ref{alternativeexpressions}). Following Ref.~\cite{coecke2013causal}, we use \trace{} 
     to denote partial trace (here slightly generalised, to include the partial trace of a wire carrying an $i$ index, defined in an obvious way.).  }
\end{figure}

Analogous circuit diagrams can be provided in the quantum case, as depicted in Fig.~\ref{QCircuits}, with analogous interpretations of the various equalities. Since quantum systems cannot be copied, however, something must replace the dot that appears in the lower two circuits of Fig.~\ref{CCircuits}. For the lower two circuits of Fig.~\ref{QCircuits}, we introduce a new symbol that indicates the decomposition of the Hilbert space $\mathcal{H}_A$ into a direct sum of tensor products, as per condition~4 of Thm~\ref{alternativeexpressions}. The symbol is a circle decorated with the set $\{ i \}$, where the value $i$ indexes the terms in the direct sum. For each value of $i$, the left-hand wire carries the factor $\mathcal{H}_{A_i^L}$ and the right-hand wire the factor $\mathcal{H}_{A_i^R}$. 

In the lower right circuit, the gates represent unitary channels, and are labelled with the corresponding unitary operators $V$ and $W$ (as opposed to the Choi-Jamio\l{}kowski channel operators). The unitary operator $V$, for example, labels a gate whose action is confined to the left-hand factors in this decomposition, along with the system $\lambda_B$. The interpretation, roughly, is that the form of $V$ must respect the decomposition of $\mathcal{H}_A$. More precisely, the unitary operator can be written as a matrix that is block diagonal with respect to the subspace decomposition, with the $i$th block being of the form $V_i \otimes I_{A_i^R}$ for a unitary matrix $V_i$ acting on $\mathcal{H}_{\lambda_B} \otimes \mathcal{H}_{A_i^L}$. Similarly, $W$ can be written as a block diagonal matrix, with the $i$th block of the form  $I_{A_i^L}  \otimes W_i$ for a unitary matrix $W_i$ acting on $\mathcal{H}_{A_i^R} \otimes \mathcal{H}_{\lambda_C}$.

In the lower left circuit, in a slight mixing of notation, gates are labelled with the channel operators $\rho_{B|A}$ and $\rho_{C|A}$ \footnote{The analogous mixed notation also appears in Fig.~\ref{CCircuits} for the classical case.}. Suppose that, as in the figure, a channel operator $\rho_{B|A}$ labels a gate whose action is confined to the left-hand factors in the decomposition, along with another system $\lambda_B$. This indicates that the channel corresponds to a set of Kraus operators $\{ K^j \}$, where for each $j$, the Kraus operator $K^j$ is block diagonal, with the $i$th block being of the form $K^j_i \otimes I_{A_i^R}$, with $K^j_i$ acting on $\mathcal{H}_{\lambda_B}\otimes \mathcal{H}_{A_i^L}$. Similarly for $\rho_{C|A}$, the right hand factors and the system $\lambda_C$.\footnote{In introducing the circle notation, we have defined the action of gates on the left/right factors in such a way that coherence between different subspaces in the direct sum can be maintained. This corresponds to the fact that the condition of decomposing into the appropriate form is applied to the unitary or Kraus operators, rather than to the channel operator itself. We have done this on the grounds that with this definition, the circle notation is most likely to be useful in future applications. In the lower two circuits of Fig.~\ref{QCircuits}, however, note that coherence between the different subspaces is lost. In the lower right circuit, coherence is lost when the partial traces are performed on the extra outgoing wires. In the lower left circuit, the final output admits a global factorization of the form $B \otimes C$, and output wires carrying an $i$ index do not even appear, indicating that this degree of freedom has been traced out. 
Each Kraus operator, in this case, must act non-trivially only on the $i$th subspace, for some $i$, and one may deduce that $\rho_{B|A}$ is of the form $\sum_i \rho_{B|A_i^L} \otimes I_{A_i^R}$, and similarly $\rho_{C|A}$, consistently with condition~4 of Thm~\ref{alternativeexpressions}.}\footnote{Clearly, the notation can be extended in various ways, to include circles with multiple output wires, circles indicating a further decomposition following another circle, and so on. A fully general interpretation and calculus for these extended circuit diagrams is left for future work.}

The equivalences of Fig.~\ref{QCircuits} can now be summarized as follows. Equality (1) simply asserts the fact that $\rho_{BC|A}$ admits a unitary dilation. Equality (4) asserts that the channel $\rho_{BC|A}$ is such that $B$ and $C$ are quantum conditionally independent given $A$, according to the definition we have proposed (Def.~\ref{QCIoutgivenin}). This equality follows from the expression for quantum conditional independence described in condition 4 of Thm~\ref{alternativeexpressions}. Equality (3) asserts that the channels $\rho_{B|A}$ and $\rho_{C|A}$ separately admit unitary dilations. Equality (2) asserts that $\rho_{BC|A}$ is compatible with $A$ being a complete common cause of $B$ and $C$ by depicting conditions under which $\lambda_B$ has no influence on $C$ and $\lambda_C$ has no influence on $B$. Here, the unitary matrix $U$ is decomposed as $U = (I_{\lambda_B}\otimes W)(V\otimes I_{\lambda_C})$, as per the proof of Thm~\ref{maintheorem}.

\subsection{Examples}\label{examples}

\subsubsection{A unitary transformation}

\begin{figure}
\includegraphics[scale=0.35,angle=0]{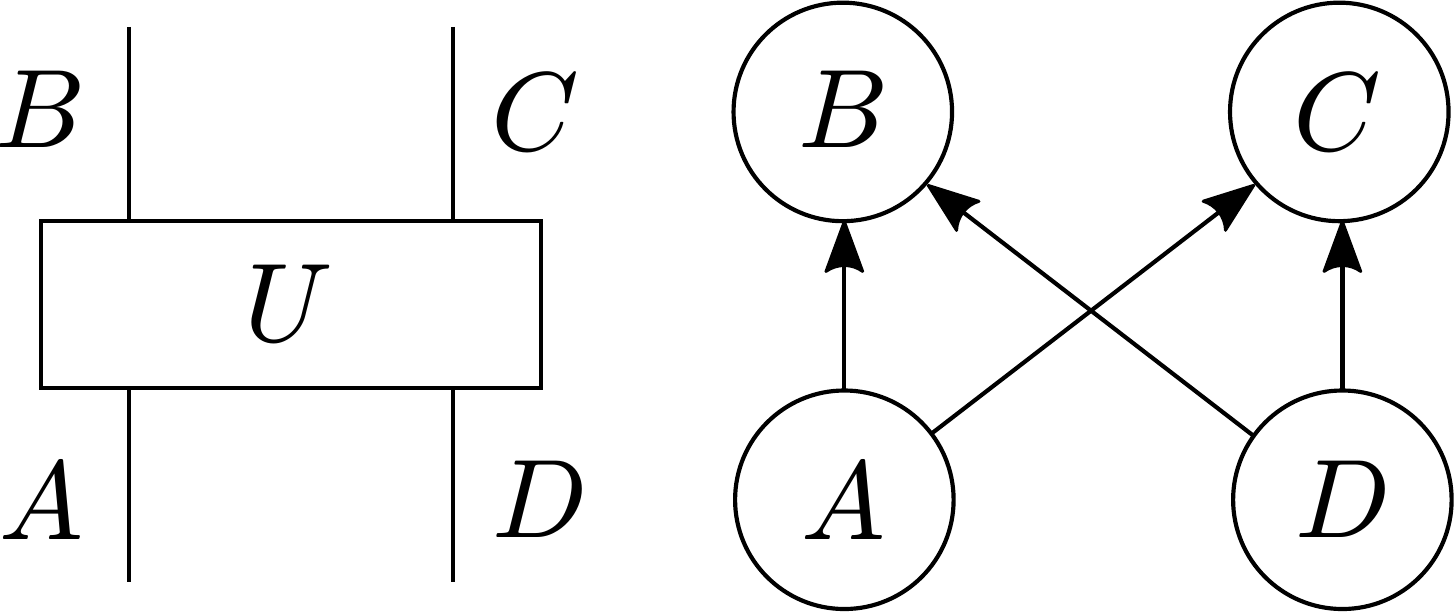}
\caption{For a generic unitary transformation from $A D$ to $BC$, the complete common cause of $B$ and $C$ is the composite system $A D$.}\label{bipartiteunitary}
\end{figure}

Consider the case in which inputs $A$ and $D$ evolve, via a generic unitary transformation $U$ into outputs $B$ and $C$. 
In Fig.~\ref{bipartiteunitary}, we illustrate the circuit and the corresponding causal diagram.

The channel $\rho_{BC|A D}$ which one obtains in this case is compatible with the complete common cause of $B$ and $C$ being the composite system $A D$.  This follows from the fact that $\rho_{BC|A D}$ has a trivial dilation, which is to say that the ancillary system is not required, and therefore 
 trivially satisfies the condition for compatibility laid out in Def.~\ref{Qcompatibility}.   It follows from Thm~\ref{maintheorem} that for such a $\rho_{BC|A D}$, the outputs $B$ and $C$ are quantum conditionally independent given the input $A D$, which means that $\rho_{BC|A D}= \rho_{B|A D}\rho_{C|A D}$, as can also be verified by direct calculation.  Similarly, the alternative expressions for this sort of quantum conditional independence, namely, conditions 3 and 4 of Thm~\ref{alternativeexpressions}, can be verified to hold.

\subsubsection{Coherent copy vs. incoherent copy}\label{coherentvsincoherent}

Consider the simple example of a classical channel, taking $X$ to $Y,Z$, where $X,Y,Z$ are bit-valued and the mapping between input strings and output strings is
\begin{align}
0_X &\rightarrow 0_Y 0_Z, \nonumber\\
1_X &\rightarrow 1_Y 1_Z.\label{classicalcopy}
\end{align}
The outputs of the channel are conditionally independent given the input;
variation in $X$ fully explains any correlation between $Y$ and $Z$. Indeed this example may be seen as the paradigmatic case of the explanation of classical correlations via a complete common cause.

One quantum analogue of this channel is the incoherent copy of a qubit: a qubit $A$ is measured in the computational basis; if $0$ is obtained, then prepare the state $|00\rangle_{BC}$ and if $1$ is obtained, prepare $|11\rangle_{BC}$. The operator representing this channel is
\[
\rho_{BC|A} =  |000\rangle\langle 000|_{BCA^*} + |111\rangle\langle 111|_{BCA^*}.
\]
It is easily verified that this operator satisfies each of the conditions of Thm~\ref{maintheorem}, so that $B$ and $C$ are quantum conditionally independent given $A$ for this channel. The decomposition of the $A$ Hilbert space implied by Condition~4 is 
\[
\mathcal{H}_A = \left( \mathbb{C} \otimes \mathbb{C} \right) \oplus \left( \mathbb{C}\otimes \mathbb{C} \right),
\]
where $\mathbb{C}$ is the $1$-dimensional complex Hilbert space, i.e., the complex numbers.

The other direct quantum analogue of the classical copy above is the channel that makes a coherent copy of a qubit, where the mapping from input states to output states is:
\begin{equation} \label{eqn:coherent-copy}
\alpha |0\rangle_A + \beta |1\rangle_A \rightarrow \alpha |0\rangle_B |0\rangle_C + \beta |1\rangle_B |1\rangle_C.
\end{equation}
This channel is represented by the operator 
\[
\rho_{BC|A} = (|000\rangle_{BCA^*} + |111\rangle_{BCA^*})(\langle 000|_{BCA^*} + \langle 111|_{BCA^*}),
\]
which corresponds to an unnormalized GHZ state. It can easily be verified that $I(B:C|A)=1$ for a trace-one version of this state, hence it is {\em not} the case that outputs $B$ and $C$ are quantum conditionally independent given the input $A$. There is, then, no way in which this channel can arise as a marginal channel in a situation in which $A$ is the complete common cause of $B$ and $C$. 

At first blush, this conclusion may seem surprising. Given the mapping described by Eq.~(\ref{eqn:coherent-copy}), where would correlations between outputs $B$ and $C$ come from, other than being completely explained by the input $A$? 

\begin{figure}
\includegraphics[scale=0.35,angle=0]{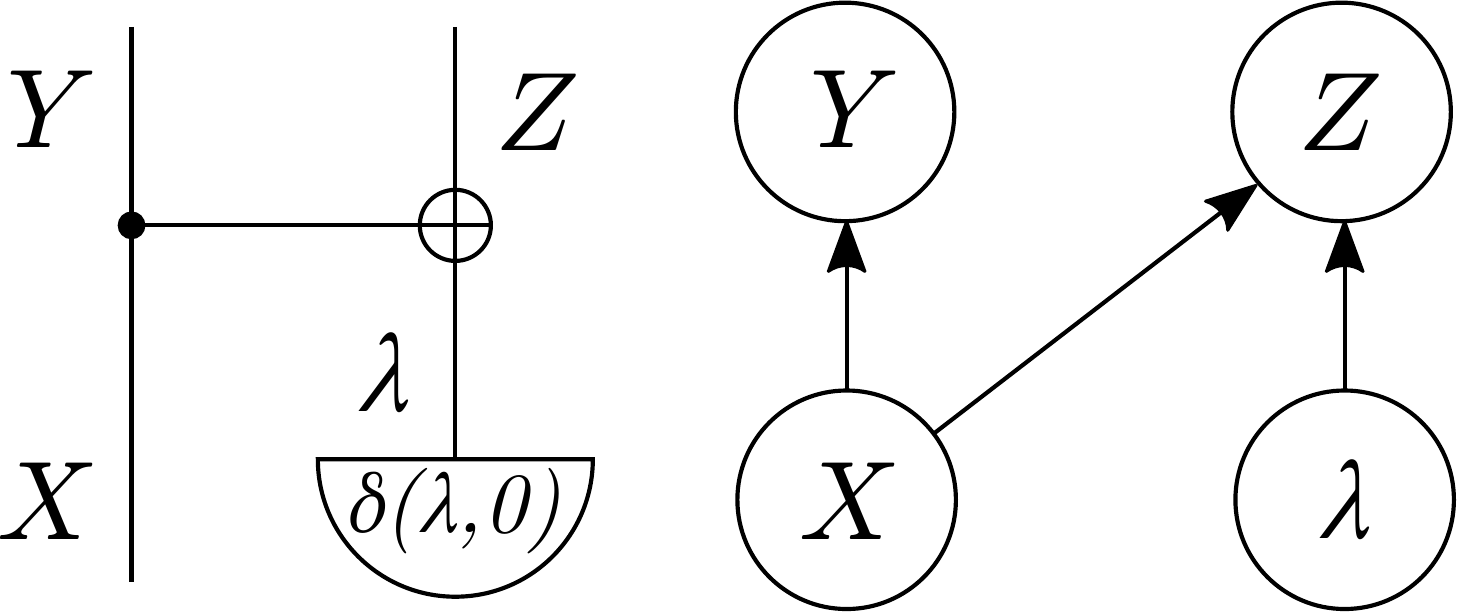}
\caption{Classical realization of a copy operation using an ancilla and classical CNOT gate.}\label{classicalcnot}
\end{figure}

\begin{figure}
\includegraphics[scale=0.35,angle=0]{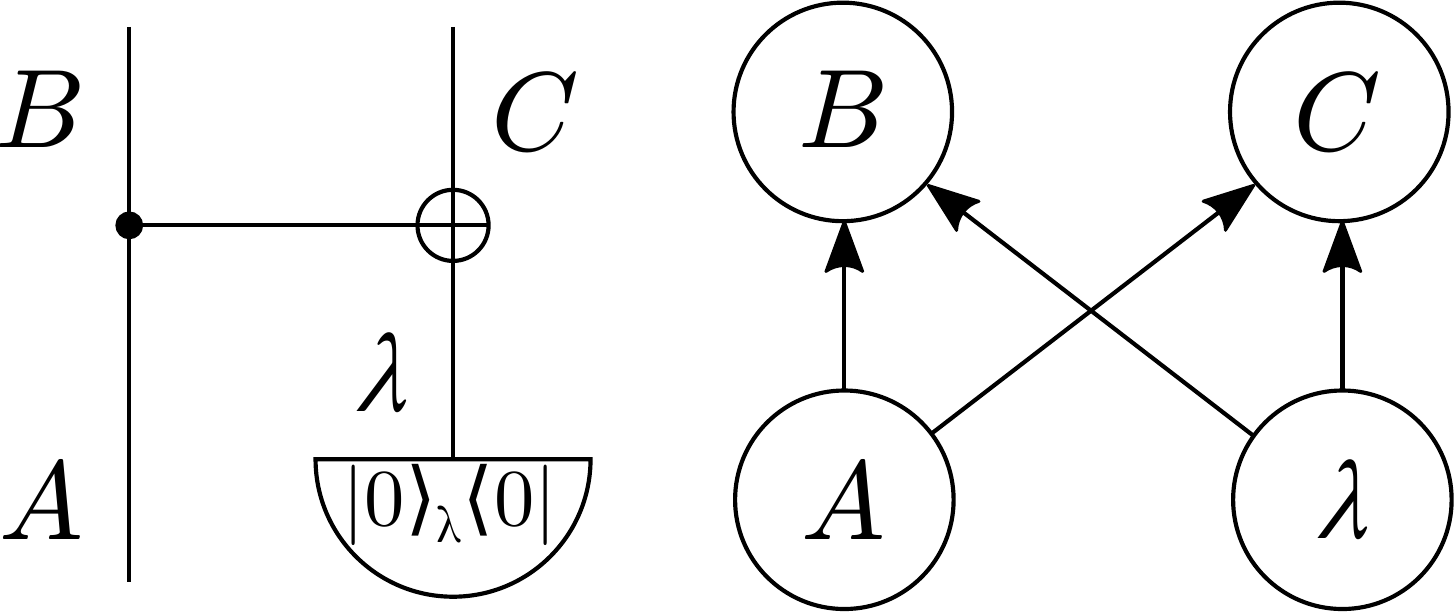}
\caption{Quantum realization of the coherent copy using an ancilla and quantum CNOT gate.}\label{quantumcnot}
\end{figure}

The puzzle is resolved by considering the dilation of the coherent copy to a unitary transformation, and the interpretation of quantum pure states. Consider Figs.~\ref{classicalcnot} and \ref{quantumcnot}, which respectively show a classical copy operation via the classical CNOT gate and a quantum coherent copy operation via the quantum CNOT gate 
 \footnote{The quantum version in Fig.~\ref{quantumcnot} was studied for similar reasons in Ref.~\cite{schumacher2012isolation}, though from a different perspective}. 

In the classical case, there are two reasons why any correlation between $Y$ and $Z$ must be entirely explained by statistical variation in the value of $X$. First, the ancillary variable $\lambda$ is prepared deterministically with value $0$, so there is no possibility that statistical variation in the value of $\lambda$ underwrites the correlations between $B$ and $C$. Second, the mapping between input strings and output strings for the classical CNOT gate,
\begin{align}
0_X 0_{\lambda} &\rightarrow 0_Y 0_Z, \nonumber\\
0_X 1_{\lambda} &\rightarrow 0_Y 1_Z, \nonumber\\
1_X 0_{\lambda} &\rightarrow 1_Y 1_Z, \nonumber\\
1_X 1_{\lambda} &\rightarrow 1_Y 0_Z,\label{eq:classicalCNOT}
\end{align}
(which one easily verifies to reduce to the classical copy of Eq.~\eqref{classicalcopy} when one sets $\lambda$ to 0),
has the causal structure depicted in Fig.~\ref{classicalcnot}, so that $\lambda$ does not act as a common cause of $Y$ and $Z$ but only a local cause of $Z$. 

In the quantum case, neither reason applies.
Concerning the second reason, the quantum CNOT has the causal structure depicted in Fig.~\ref{quantumcnot}:
the quantum CNOT is such that not only does $A$ have a causal influence on $C$, but $\lambda$ has a causal influence on $B$ as well.  In other words, unlike the classical CNOT, there is a back action of the target on the control. 
It follows that in the quantum case, $\lambda$ {\em can} act as a common cause of $B$ and $C$.  
Furthermore, the ancilla is prepared in a quantum pure state $|0\rangle$. This is dis-analogous to a point distribution on the value 0 for the classical variable $\lambda$ if one takes the view that a quantum pure state represents \emph{maximal but incomplete} information about a quantum system~\cite{CFS02arxiv,fuchs2002quantum,Spekkens2007, Leifer-06, spekkens2016quasi}.  In this case, one must allow for the possibility that some correlation between $B$ and $C$ is 
due to the ancilla, in which case $A$ is not the complete common cause of $B$ and $C$~
\footnote{It is interesting to consider an exactly analogous scenario, as it arises in the toy theory of Ref.~\cite{Spekkens2007}. Here, a system analogous to a qubit can exist in one of four distinct classical states (the \emph{ontic} states of the system). But an agent who prepares systems and measures them can only ever have partial information about which of the four ontic states a system is in. The toy equivalent of a CNOT gate corresponds to a reversible deterministic map, i.e., a permutation of the ontic states. By considering the probability distribution over ontic states of the various systems, one may verify directly that the ontic states of toy systems $B$ and $C$ are not
determined by
the ontic state of toy system $A$. Rather, the ontic states of $B$ and $C$ depend also on the ontic state of $\lambda$.  Furthermore, the analogue of a pure quantum state for $\lambda$ is a probability distribution on $\lambda$ that is not a point distribution.  In this way, statistical correlations between $B$ and $C$ can be underwritten by statistical variation in the ontic state of $\lambda$. 
}.

\subsection{Generalization to one input, $k$ outputs}

Thms~\ref{maintheorem} and \ref{alternativeexpressions}, which apply to quantum channels with one input and two outputs, can be generalized to the case of one input and $k$ outputs.  

Consider a channel $\rho_{B_1\ldots B_k | A}$, and let $\bar{B}_i$ denote the collection of all outputs apart from $B_i$.
The notion of quantum compatibility from Def.~\ref{Qcompatibility} generalizes in the obvious way:
$\rho_{B_1\ldots B_k|A}$  is said to be compatible with $A$ being a complete common cause of $B_1\ldots B_k$, if it is possible to find ancillary quantum systems $\lambda_1,\ldots , \lambda_k$, states $\rho_{\lambda_1},\ldots, \rho_{\lambda_k}$, and a unitary channel 
 where, for each $i$, $\lambda_i$ has no causal influence on $\bar{B}_i$, such that these constitute a dilation of $\rho_{B_1\ldots B_k|A}$.

The generalization of Thms~\ref{maintheorem} and \ref{alternativeexpressions}, consolidated into a single theorem, is  as follows:
\begin{theorem}\label{maintheoremmany}
 The following are equivalent:
\begin{enumerate}
\item  $\rho_{B_1\ldots B_k|A}$ is compatible with $A$ being a complete common cause of $B_1\ldots B_k$. \color{black}
\item $\rho_{B_1\ldots B_k | A} = \rho_{B_1|A} \cdots \rho_{B_k|A}$, where for all $i$, $j$, $[\rho_{B_i|A}, \rho_{B_j|A}]=0$.
\item For each $i$, $I(B_i : \bar{B}_i | A) = 0$ where $I(B_i : \bar{B}_i | A)$ is the quantum conditional mutual information evaluated on the (positive, trace-one) operator $\hat{\rho}_{B_1\ldots B_k|A}$. \color{black}
\item The Hilbert space for the $A$ system can be decomposed as $\mathcal{H}_{A} = \bigoplus_i \mathcal{H}_{A_i^1}\otimes\cdots\otimes\mathcal{H}_{A_i^k}$ such that $\rho_{B_1\ldots B_k|A} = \sum_i \left(\rho_{B_1|A_i^1}\otimes\cdots\otimes\rho_{B_k|A_i^k}\right)$, where for each $i$, and each $l\in \{1,\dots,k\}$, $\rho_{B|A_i^l}$ represents a CPTP map $\mathcal{B}(\mathcal{H}_{A_i^l}) \rightarrow \mathcal{B}(\mathcal{H}_{B_l})$.
\end{enumerate}
\end{theorem}
\color{black}
The proof is in Appendix~\ref{maintheoremmanyproof}. 
By analogy to the classical case, if conditions 2, 3 and 4 of Thm~\ref{maintheoremmany} hold, we say that $B_1\ldots B_k$ are quantum conditionally independent given $A$ for the channel $\rho_{B_1\ldots B_k|A}$. 

\section{Classical causal models}\label{ccausalmodels}

\subsection{Definitions}

Reichenbach's principle is important because it generalizes to the modern formalism of \emph{causal models} \cite{Pearl-09,Spirtes-Glymour-01}.

A causal model consists of two entities: (i) a causal structure, represented by a directed acyclic graph (DAG) where the nodes represent random variables and the directed edges represent the directed causal influences among these (several examples have already been presented in this article), and (ii) some parameters, which specify the strength of the causal dependences and the probability distributions for the variables associated to root nodes in the DAG (i.e., those with no incoming arrows). Some terminology is required to present the formal definitions.

 Given a DAG with nodes $X_1,\ldots, X_n$, let $\mathrm{Parents}(i)$ denote the parents of node $X_i$, that is, the set of nodes that have an arrow into $X_i$, and let $\mathrm{Children}(i)$ denote the children of node $X_i$, that is, the set of nodes $X_j$ such that there is an arrow from $X_i$ to $X_j$. The descendents of $X_i$ are those nodes $X_j$, $j\ne i$, such that there is a directed path from $X_i$ to $X_j$. The ancestors of $X_i$ are those nodes $X_j$ such that $X_i$ is a descendent of $X_j$.

\begin{definition}\label{classicalcausalnetworkdef}
A \emph{causal model} specifies a DAG, with nodes corresponding to random variables $X_1,\ldots, X_n$, and a family of conditional probability distributions $\{ P(X_i | \mathrm{Parents}(i))\}$, one for each $i$. 
\end{definition}

\begin{definition}\label{Markov property}
Given a DAG, with random variables $X_1,\ldots, X_n$ for nodes, and given an arbitrary joint distribution $P(X_1\ldots X_n)$, the distribution is said to be \emph{Markov for the graph} if and only if it can be written in the form of 
\begin{equation}\label{Markov}
P(X_1\ldots X_n) = \prod_{i=1}^n P(X_i | \mathrm{Parents}(i)).
\end{equation} 
(Recall that each conditional $P(X_i | \mathrm{Parents}(i))$ can be computed from the joint $P(X_1\ldots X_n).$)
\end{definition}

The generalization of Reichenbach's principle that is afforded by the formalism of causal models is this:
if there are statistical dependences among variables $X_1,\ldots, X_n$, expressed in the particular form of the joint distribution  $P(X_1\ldots X_n)$, then there should be a causal explanation of these dependences in terms of a DAG relative to which the distribution  $P(X_1\ldots X_n)$ is Markov.\color{black}

Note that an alternative way of formalizing the Markov property is that $P(X_1\ldots X_n)$ is Markov for the graph if and only if, for each $i$, $P(X_i | \mathrm{Parents}(i)) = P(X_i | \mathrm{Nondesc}(i))$, where $\mathrm{Nondesc}(i)$ is the set of non-descendents of node $X_i$.  The intuitive idea is that the parents of a node screen off that node from the other nondescendents: once the values of the parents are fixed, the values of other non-descendent nodes are irrelevant to the value of $X_i$.

Note also that 
Reichenbach's principle is easily seen to be a special case of the requirement that for a joint  distribution to be explainable by the causal structure of some DAG, it must be Markov for that DAG:  
if two variables, $Y$ and $Z$, are ancestrally independent in the graph, then any distribution that is Markov for this graph must factorize on these, $P(Y Z) = P(Y)P(Z)$, which is the qualitative part of Reichenbach's principle in its contrapositive form;  
if two variables, $Y$ and $Z$, have a variable $X$ as a complete common cause, as in the DAG of Fig.~\ref{xtoyz}, then any distribution that is Markov for the graph must satisfy $P(YZ|X) = P(Y|X)P(Z|X)$, which is the quantitative part of Reichenbach's principle.

\subsection{Justifying the Markov condition}

Just as we previously asked whether there was some principle that forced a rational agent to assign probability distributions in accordance with the quantitative part of Reichenbach's principle, we can similarly ask why a rational agent who takes causal relationships to be given by a particular DAG should arrange their beliefs so that the joint distribution is Markov for the DAG. 

The justification of the Markov condition parallels the justification of the quantitative part of Reichenbach's principle that was presented in Sec.~\ref{justifyingclassicalreichenbach}. We begin by outlining what the qualitative part of Reichenbach's principle and the assumption of fundamental determinism imply for any arbitrary causal structure.

\begin{definition}[Classical compatibility with a DAG] $P(X_1\ldots X_n)$ is said to be compatible with a DAG $G$ with nodes $X_1,\ldots, X_n$ if one can find a DAG $G'$  that is obtained from $G$ by adding extra root nodes $\lambda_1,\ldots,\lambda_n$, such that for each $i$, the node $\lambda_i$ has a single outgoing arrow, to $X_i$, 
and one can find, for each $i$, a distribution $P(\lambda_i)$ and a function $f_i$ from $(\lambda_i, \mathrm{Parents(i)})$ to $X_i$  such that  
\begin{align}
&P(X_1\ldots X_n)\nonumber\\
& = \sum_{\lambda_1\ldots \lambda_n} \left[ \prod_{i=1}^n \delta (X_i , f_i(\lambda_i, \mathrm{Parents}(i))) P(\lambda_i)\right].\nonumber
\end{align} 
\end{definition}

\begin{theorem}[Ref.~\cite{Pearl-09}]\label{expand}
Given a joint distribution $P(X_1\ldots X_n)$ and a DAG $G$ with nodes $X_1,\ldots, X_n$, the following are equivalent:
\begin{enumerate}
\item $P(X_1\ldots X_n)$ is compatible with the causal structure described by the DAG $G$.
 \item $P(X_1\ldots X_n)$ is Markov for $G$, that is, 
 \[
P(X_1\ldots X_n) = \prod_{i=1}^n P(X_i | \mathrm{Parents}(i)).
\]
\end{enumerate}
\end{theorem}

The $1 \to 2$ implication in Thm~\ref{expand} can be read as follows: if it is granted that causal relationships are indicative of underlying deterministic dynamics, and that the qualitative part of Reichenbach's principle is valid,
 then, on pain of irrationality, an agent's assignment $P(X_1\ldots X_n)$ must be Markov for the original graph. 

The $2 \to 1$ implication in Thm~\ref{expand}, like that of Thm~\ref{Cversionmaintheorem}, is pertinent for causal inference.  It asserts that if one observes a distribution $P(X_1\ldots X_n)$, then of the causal models that are compatible with this distribution, the only ones that do not require fine-tuning of the parameters 
are those involving DAGs relative to which the distribution is Markov.
\color{black}

\section{Quantum causal models}
\label{qcausalnetworks}

\subsection{The proposed definition}

In our treatment of the simple causal scenario where $A$ is a complete common cause of $B$ and $C$ (the DAG of Fig.~\ref{atobc}), we focussed on what form is implied for the quantum channel $\rho_{BC|A}$.
But there has not been any attempt to define a quantity analogous to the classical joint distribution, that is, a quantity analogous to $P(XYZ)$ in the case of the DAG of Fig.~\ref{xtoyz}, nor indeed other classical Bayesian conditionals such as $P(X|YZ)$. For works that aim to achieve such analogues, 
see Ref.~\cite{Leifer-06,Leifer-Spekkens-13}. See also Ref.~\cite{horsman2016can}, however, 
where it is shown that if one associates a single Hilbert space to a system at a given time, then there are significant obstacles to establishing an analogue of a classical joint distribution when the set of quantum systems includes some that are causal descendants of others

This work takes a different approach. The interpretation of a quantum causal model will be that each node represents a local region of time and space, with channels such as $\rho_{BC|A}$ describing the evolution of quantum systems in between these regions. At each node, there is the possibility that an agent is present with the ability to intervene inside that local region. Each node $A_i$ will then be associated with two Hilbert spaces, one corresponding to the incoming system (before the agent's intervention) and the dual space, which corresponds to the outgoing system (after the agent's intervention). A quantum causal model will consist of a specification, for every node, of the quantum channel from its parents to the node,
with the operational significance of a network being that it is used to calculate joint probabilities for the agents to obtain the various possible joint outcomes for their interventions. This way of treating quantum systems over time has appeared in various different approaches in the literature, including the multi-time formalism 
\cite{aharonov2007two,AharonovEtAl_2009,aharonov2013each,silva2014pre}, the quantum combs formalism
\cite{ChiribellaEtAl_2009,Chiribella_2012,ChiribellaEtAl_2013}, the process matrices formalism
 \cite{Oreshkov2012,AraujoEtAl_2014,Costa2016}, and a number of other works as well \cite{Oeckl_2003,oreshkov2014operational,oreshkov2015operational,Ried-Agnew-15}.

The discussion of classical causal models in Sec.~\ref{ccausalmodels}, and the results of Sec.~\ref{qReichenbach} for the special case of $A$ a complete common cause of $B$ and $C$, suggest the following generalization. 
\begin{definition}\label{quantumnetworkdef}
A \emph{quantum causal model} specifies a DAG, with nodes $A_1,\ldots, A_n$, supplemented with the following. For each node $A_i$, there is associated a finite-dimensional Hilbert space $\mathcal{H}_i$ (the `input' Hilbert space), and the dual space $\mathcal{H}^\ast_i$ (the `output' Hilbert space). 
For each node $A_i$, there is associated a quantum channel, described by an operator $\rho_{A_i | \mathrm{Parents}(i)} \in \mathcal{B}(\mathcal{H}_i \otimes \mathcal{H}_{\mathrm{Parents}(i)}^\ast ) $, where $\mathcal{H}_{\mathrm{Parents}(i)}^\ast$ is the tensor product of the output Hilbert spaces associated with the parents of $A_i$. These channels commute pairwise, i.e., for any $i,j$, $[\rho_{A_i | \mathrm{Parents}(i)}, \rho_{A_j | \mathrm{Parents}(j)}]=0$ (which is a nontrivial constraint whenever $\mathrm{Parents}(i) \cap \mathrm{Parents}(j)$ is nonempty). The overall state is respresented by an operator on $\bigotimes_{i=1}^n \mathcal{H}_{{\bf A}_i}$, where $\mathcal{H}_{{\bf A}_i} := \mathcal{H}_{A_i} \otimes \mathcal{H}_{A_i}^\ast$, denoted $\sigma_{{\bf A}_1\ldots {\bf A}_n}$ and given by
\begin{equation}\label{quantummarkov}
\sigma_{{\bf A}_1\ldots {\bf A}_n} = \prod_{i=1}^n \rho_{A_i|\mathrm{Parents}(i)}.
\end{equation}
\end{definition}

Recall from Section~\ref{qReichenbach} that, given a quantum channel $\rho_{BC|A}$, it is compatible with $A$ being the complete common cause of $B$ and $C$ if and only if 
$\rho_{BC|A} = \rho_{B|A}\rho_{C|A}$, and if this holds, then $[\rho_{B|A},\rho_{C|A}]=0$. The definition of a quantum causal model, in particular, the stipulation that the channels commute pairwise, generalizes this idea.

\subsection{Making predictions}

In order to see how a quantum causal model is used to calculate probabilities for the outcomes of agents' interventions, consider a quantum causal model with nodes $A_1,\ldots ,A_n$ and state $\sigma_{{\bf A}_1 \ldots {\bf A}_n}$. 
Let the intervention at node $A_i$ have classical outcomes labelled by $k_i$. The intervention is defined by a quantum instrument (that is, by a set of completely-positive trace-non-increasing maps, one for each outcome) which sum to a trace-preserving map. In order to write the probabilities for the outcomes in a simple form, it is useful to define the instrument in such a way that the map associated to each outcome takes operators on $\mathcal{H}_{A_i}^\ast$ into operators on $\mathcal{H}_{A_i}^\ast$. Hence, suppose that the outcome $k_i$ corresponds to the map $\mathcal{E}_{A_i}^{k_i} : \mathcal{B}(\mathcal{H}_{A_i}^\ast)\rightarrow \mathcal{B}(\mathcal{H}_{A_i}^\ast)$ and let
\[
\tau_{{\bf A}_i}^{k_i} = \sum_{lm} \mathcal{E}_{A_i}^{k_i} (|l\rangle_{A_i^\ast}\langle m|) \otimes |l\rangle_{A_i}\langle m|.
\]
The outcome $k_i$ of the agent's intervention can then be represented by the (positive, basis-independent) operator $\tau_{{\bf A}_i}^{k_i}$ isomorphic to $\mathcal{E}_{A_i}^{k_i}$. 

If an agent does not intervene at the node $A_i$, this corresponds to the linking operator itself,
\[
\tau_{{\bf A}_i}^{\rm id} = \sum_{lm} |l\rangle_{A_i^\ast}\langle m| \otimes |l\rangle_{A_i}\langle m|.
\]
The joint probability for the agents to obtain outcomes $k_1,\ldots , k_n$ is given by
\begin{align}\label{Qrecords}
P(k_1\ldots k_n) = \mathrm{Tr} (\sigma_{{\bf A}_1\ldots {\bf A}_n} (\tau_{{\bf A}_1}^{k_1}\otimes \cdots \otimes \tau_{{\bf A}_n}^{k_n})).
\end{align}

We can also define operations on the state $\sigma_{{\bf A}_1\ldots {\bf A}_n}$ corresponding to marginalization over the outcome $k_i$ of an intervention on node $A_i$ by 
$\sum_{k_i} \mathrm{Tr}_{{\bf A}_i }$. 
In this case, the joint state on the rest of the nodes after such marginalization is
\[
\sigma_{{\bf A}_1\ldots {\bf A}_{(i-1)} {\bf A}_{(i+1)}\ldots {\bf A}_n} = \sum_{k_i} \mathrm{Tr}_{{\bf A}_i} (\sigma_{{\bf A}_1\ldots {\bf A}_n} \tau_{{\bf A}_i}^{k_i}  ).
\]
If the intervention at node $A_i$ is trivial, then 
\[
\sigma_{{\bf A}_1\ldots {\bf A}_{(i-1)} {\bf A}_{(i+1)}\ldots {\bf A}_n} = \mathrm{Tr}_{{\bf A}_i} (\sigma_{{\bf A}_1\ldots {\bf A}_n} \tau_{{\bf A}_i}^{\rm id}  ).
\]

\subsection{Classical interventional models}

Given the proposed definition of a quantum causal model, and the interpretation in terms of agents intervening at nodes, there is a stronger analogy to be made with a classical formalism that similarly involves interventions, than there is to the standard classical causal models introduced in Sec.~\ref{ccausalmodels}. 

In order to make this explicit, consider a classical interventional causal model constructed as follows. For a given DAG, split every node $X_i$ into a pair of disconnected nodes, denoted $X^O_i$ and $X^I_i$, such that in the DAG that results, $X^I_i$ has as parents the set of nodes $\mathrm{Parents}^O(i) := \{ X^O_j : X_j \in \mathrm{Parents}(i) \}$, and $X^O_i$ has as children $\{ X^I_j : X_j \in \mathrm{Chidren}(i) \}$. In other words, the `$I$' version of each node $X_i$ has as parents the `$O$' version of each node that was a parent of $X_i$ in the original graph, and the `$O$' version of each node $X_i$ has as children the `$I$' version of each node that was a child of $X_i$ in the original graph. In this case, one can represent the resulting DAG by a conditional probability distribution
\begin{align}\label{splitnodemodel}
P(X^I_1 \ldots X^I_n | X^O_1 \ldots X^O_n ) = \prod_{i=1}^{n} P(X^I_i | \mathrm{Parents}^O(i)).
\end{align}
Our association of each node $A_i$ of the DAG  with a pair of Hilbert spaces, $\mathcal{H}_{A_i}$ and $\mathcal{H}^{*}_{A_i}$, is simply a quantum version of the splitting of a classical variable $X_i$ into $X^O_i$ and $X^I_i$, and our joint state $\sigma_{ {\bf A}_1 \ldots {\bf A}_n}$ is the quantum analogue of the conditional probability $P(X^I_1 \ldots X^I_n | X^O_1 \ldots X^O_n )$. 

In a classical interventional causal model, one can imagine an intervention at node $X_i$ as a causal process acting between $X^I_i$ and $X^O_i$ and possibly outputing an additional classical variable $k_i$ which acts as a record of some aspect of the intervention.  The most general such intervention is described by a conditional probability distribution $P(k_i, X^O_i |X^I_i)$ \footnote{Classical interventional models, as we have described them, seem rarely to be studied in full generality in the classical literature. However, among the possible intervention schemes are included the following special cases: ignoring $X^I_i$ and repreparing $X^O_i$ with a fixed value $x$ of which one keeps a record, corresponding to $P(k_i, X^O_i |X^I_i)= \delta_{k_i,x} \delta_{X^O_i,x}$ (the standard notion of intervention, as set out e.g., in~\cite{Pearl-09}); ignoring $X^I_i$ and repreparing $X^O_i$ with a value that is sampled randomly and independently of $X^I_i$ and of which one keeps a record, corresponding to $P(k_i, X^O_i |X^I_i)=  \delta_{k_i, X^O_i} P( X^O_i)$ (a randomized trial); observing $X^I_i$, keeping a record of this value and preparing $X^O_i$ to have this value, corresponding to  $P(k_i, X^O_i |X^I_i)=  \delta_{k_i, X^I_i} \delta_{X^O_i, X^I_i}$ (passive observation of $X_i$) ; observing $X^I_i$ and keeping a record of its value and repreparing $X^O_i$ to have a fixed value $x$, corresponding to $P(k_i, X^O_i |X^I_i)=  \delta_{k_i, X^I_i} \delta_{X^O_i, x} $ (the sort of intervention considered in single-world intervention graphs~\cite{richardson2013single});  simply letting the value of $X^O_i$ track the value of $X^I_i$, corresponding to $k_i$ being trivial, and $P(X^O_i |X^I_i)=  \delta_{X^O_i, X^I_i}$ (no observation being made); and many others besides.}.  
After specifying the nature of the intervention at each node, $\{ P(k_i, X^O_i |X^I_i) \}_i$, one can compute the joint probability distribution over the record variables to be
\begin{align}\label{distrecords}
P(k_1 \ldots k_n )&=\sum_{X^I_1, X^O_1 \ldots X^I_n, X^O_n} P(X^I_1 \ldots X^I_n | X^O_1 \ldots X^O_n )\nonumber\\
& \times \prod_{i=1}^n P(k_i, X^O_i |X^I_i).
\end{align}
Clearly, our intervention operators $ \tau_{{\bf A}_i}^{k_i} $ are the quantum analogue of the intervention conditionals $P(k_i,X^O_i |X^I_i)$, and our Eq.~\eqref{Qrecords} is the quantum analogue of Eq.~\eqref{distrecords}.

\subsection{Examples}

\subsubsection{Confounding common cause}

\begin{figure}
\includegraphics[scale=0.35,angle=0]{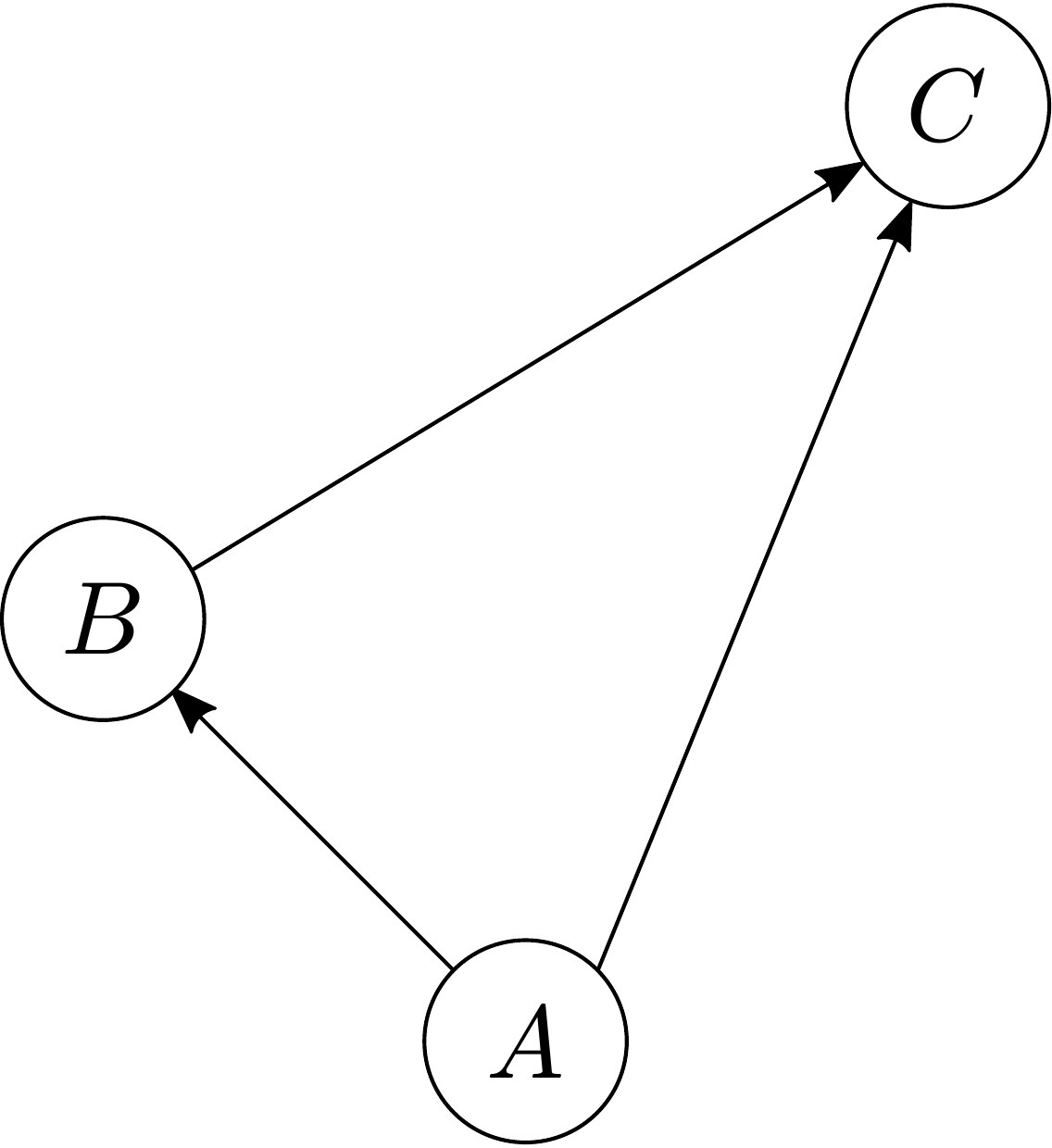}
\caption{A causal network with $A$ a common cause for $B$ and $C$ and with $B$ a parent of $C$.}\label{quantumconfound}
\end{figure}

Consider a quantum causal model with the DAG depicted in Fig.~\ref{quantumconfound}. 
The DAG is supplemented with the quantum channels $\rho_{C|AB}$, $\rho_{B|A}$, and $\rho_A$, where the latter is simply a quantum state on $\mathcal{H}_A$ (which can also be thought of as a channel from the trivial, or $1$-dimensional system into $A$). 

The corresponding state is 
$\sigma_{ABC} = \rho_{C|BA}\rho_{B|A}\rho_A$, where $\sigma$ acts on the Hilbert space  $\mathcal{H}_C^\ast\otimes \mathcal{H}_C\otimes \mathcal{H}_B^\ast\otimes \mathcal{H}_B\otimes \mathcal{H}_A^\ast\otimes \mathcal{H}_A$. By stipulation, the channels commute pairwise. This is immediate in the case of, say, $\rho_{B|A}$ and $\rho_A$, since these operators are non-trivial on distinct Hilbert spaces. But it is significant in the case of $\rho_{C|BA}$ and $\rho_{B|A}$, both of which act non-trivially on $\mathcal{H}_A^\ast$. From Thm~\ref{maintheorem}, this implies that $\mathcal{H}_A^\ast$ decomposes as $\mathcal{H}_A^\ast = \bigoplus_i \mathcal{H}_{A_i^L}^\ast\otimes \mathcal{H}_{A_i^R}^\ast$, with $\rho_{C|BA}$ acting trivially on (say) the right-hand factors and $\rho_{B|A}$ acting trivially on the left-hand factors.

The fact that the output Hilbert space of the $A$ system decomposes in this manner is a significant constraint on the kinds of quantum evolution that can be compatible with the DAG of Fig.~\ref{quantumconfound}. In words, the evolution undergone by the system emerging from $A$ is as follows: a (possibly degenerate) von Neumann measurement is performed and, controlled on the outcome, the $A$ system is split into two pieces. One piece evolves to become the input at $B$. The output at $B$ is then recombined with the other piece, and evolves to become the input at $C$.

\begin{figure}
\includegraphics[scale=0.35,angle=0]{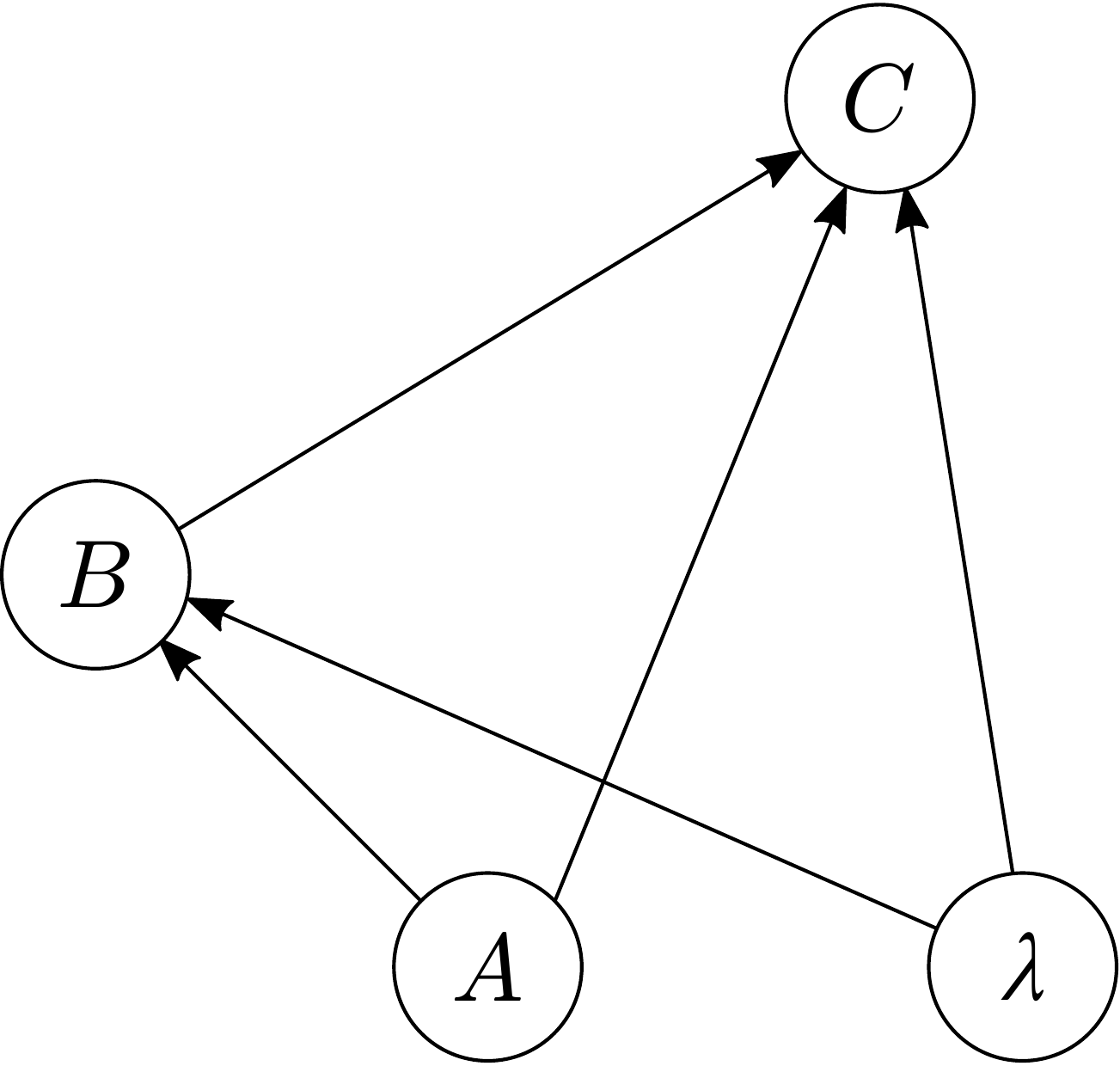}
\caption{The causal structure of Fig.~\ref{quantumconfound} with an extra node $\lambda$ which is a common cause for $B$ and $C$. A causal model with this DAG may describe a qubit interacting with an environment: $A$, $B$, $C$ represent the qubit system at three different times and $\lambda$ the environment at the initial time.}\label{quantumconfoundwithenv}
\end{figure}

By way of contrast, it is also instructive to consider quantum causal models with the causal structure shown in Fig.~\ref{quantumconfoundwithenv}. 
Such a quantum causal model may represent, for example, the non-Markovian evolution of a qubit over three time steps, with $A$, $B$ and $C$ representing the qubit at each time step, and where the qubit interacts with an environment whose initial state is $\rho_{\lambda}$. The qubit is initially uncorrelated with the environment. Suppose that the state of the environment at the second and third time steps is not of interest, hence corresponding nodes do not appear in the DAG. Given that over the course of this evolution, information can flow from the qubit to the environment, and back again, it is necessary to include an arrow from $A$ to $C$, as well as from $\lambda$ to $B$ and $\lambda$ to $C$.

A quantum causal model with this DAG defines commuting channels $\rho_{C|BA\lambda}$, $\rho_{B|A\lambda}$, $\rho_A$, $\rho_{\lambda}$. From the fact that  $\rho_{C|BA\lambda}$ and $\rho_{B|A\lambda}$ commute, we conclude that the Hilbert space $\mathcal{H}_A^\ast \otimes \mathcal{H}_{\lambda}^\ast$ decomposes as a direct sum over direct products. However, a decomposition of $\mathcal{H}_A^\ast \otimes \mathcal{H}_{\lambda}^\ast$ as a direct sum over direct products does not imply a decomposition of the Hilbert space $\mathcal{H}_A^\ast$ alone as a direct sum over direct products. Hence the evolution of the qubit is not strongly constrained as it was in the previous example. Physically, this is important: if the qubit, for example, is interacting only weakly with the environment, then its evolution certainly could not be paraphrased in terms of a strong von Neumann measurement, as it was for evolutions compatible with Fig.~\ref{quantumconfound}. 

One further remark concerning this example will help to illustrate a distinction between quantum and classical causal models. Suppose that $\rho_{\lambda}$ is the pure state $|0\rangle\langle 0|$ and that we marginalize over $\lambda$ under the assumption that an agent at the $\lambda$ node does not intervene. In classical causal models, if a root note has a point distribution, then marginalizing over that node yields a distribution over the remaining variables that is compatible with the DAG obtained by removing that node and its outgoing arrows. This does not hold in the quantum case: even for $\rho_{\lambda}$ a pure state, marginalizing over the $\lambda$ node (assuming no intervention there) in general yields an operator $\sigma_{ABC}$ that is {\em not} compatible with the DAG obtained by removing $\lambda$ (Fig.~\ref{quantumconfound}). As with the example of the coherent copy in Sec.~\ref{coherentvsincoherent}, this makes intuitive sense if one takes the view that a quantum pure state represents maximal but incomplete information. Incomplete information about the $\lambda$ system may underwrite correlations between $B$ and $C$, so that such correlations cannot be attributed entirely to system $A$ as Fig.~\ref{quantumconfound} requires. Hence, even for the environment initially in a pure state, the non-Markovian evolution of a qubit need not obey the strong constraint implied by the causal structure of Fig.~\ref{quantumconfound}. 

\subsubsection{A simple case of Bayesian updating}\label{bayesianupdating}

This section discusses another sense in which the quantum notion of conditional independence of the outputs of a channel given the input mirrors qualitatively an important aspect of the classical case.

Consider a classical causal model with the DAG of Fig.~\ref{xtoyz} and distribution $P(XYZ)$ such that $P(YZ|X) = P(Y|X)P(Z|X)$. A particular feature of this causal scenario is that if new information is obtained about the variable $Y$, for example, if an agent learns that the value of $Y$ is $y$, then the process of Bayesian updating proceeds as follows. First, update the distribution over $X$ by applying the rule
\[
\tilde{P}(X) := P(X|Y=y) = \frac{P(Y=y|X)P(X)}{P(Y=y)}.
\]
Then use the new probability distribution on $X$, $\tilde{P}(X)$, to get an updated distribution for $Z$:
\begin{equation}\label{followarrows}
P(Z|Y=y) = \sum_x P(Z|X)\tilde{P}(X),
\end{equation}
where the sum ranges over the values that $X$ may take. Roughly speaking, the process of Bayesian updating ``follows the arrows'' of the graph. For this it is crucial that the joint distribution $P(XYZ)$ satisfies $P(YZ|X) = P(Y|X)P(Z|X)$, otherwise the term $P(Z|X)$ in Eq.~(\ref{followarrows}) would have to be replaced with $P(Z|X,Y=y)$.

Consider now a quantum causal model, with the DAG of Fig.~\ref{atobc} and with state $\sigma_{ABC} = \rho_{B|A}\rho_{C|A}\rho_A$. Suppose that an agent at $B$ intervenes, obtaining outcome $k_B$, corresponding to the operator $\tau_{\bf B}^{k_B}$. The agent wishes to calculate the probability that an intervention at $C$ yields outcome $k_C$ corresponding to $\tau_{\bf C}^{k_C}$, conditioned on having obtained the outcome $k_B$, and assuming that there is no intervention at $A$. This can be done as follows. First, update the state assigned to $A$ given the knowledge of $k_B$ to 
\[
\tilde{\sigma}_{\bf A} := \sigma_{{\bf A} | k_B} = \frac{\mathrm{Tr}_{\bf B} (\sigma_{\bf AB} \tau_{\bf B}^{k_B})}{\mathrm{Tr}(\sigma_{\bf AB} (\tau_{\bf A}^{\rm id} \otimes \tau_{\bf B}^{k_B}))}.
\]
Then apply the channel $\rho_{C|A}$ to $\tilde{\sigma}_{\bf A}$ to get the state assigned to $C$ given the knowledge of $k_B$:
\[
\sigma_{{\bf C} | k_B} = \mathrm{Tr}_{\bf A}(\rho_{C|A} \tilde{\sigma}_{\bf A} \tau_{\bf A}^{\rm id}).
\]
Finally, calculate the probability of $k_C$:
\[
P(k_C | k_B) = \mathrm{Tr} (\sigma_{{\bf C} | k_B} \tau_{\bf C}^{k_C}).
\] 
Again, the process of Bayesian updating ``follows the arrows'' of the graph. Note that for this to work, it was crucial that the channel $\rho_{BC|A}$ satisfied $\rho_{BC|A} = \rho_{B|A}\rho_{C|A}$.

\section{Relation to prior work} \label{priorwork}

We now present a short review of prior works on quantum causal models and describe how our own proposal relates to these.

Preliminary work in this area took to form of explorations of Bell-type inequalities (and whether they admit of quantum violations) for novel causal scenarios~\cite{branciard2010characterizing,Fritz-12}.  Several researchers recognized that the formalism of classical causal models could provide a unifying framework in which to pose the problem of deriving Bell-type constraints, and that this framework might be extended to address the problem of deriving constraints on the correlations that can be obtained with quantum resources~\cite{Wood-Spekkens-15,chaves2014causal,Chaves-Majenz-15,Chaves-Kueng-15}. Note that such constraints are expressed entirely in terms of classical settings and classical outcomes of measurements.  Henson, Lal and Pusey~\cite{Henson-Lal-14} and Fritz~\cite{fritz-14} independently proposed definitions of quantum causal models 
with the purpose of expressing such constraints.  In these approaches, each node of the DAG represents a process (which may have a classical outcome), while each directed edge is associated with a system that is passed between processes. However, despite the fact that their frameworks incorporate the possibility of post-classical resources, they do not have sufficient structure to define conditional independences between quantum systems. 

Operational reformulations of quantum theory such as Refs.~\cite{hardy:five,Barrett:05c,AbramskyCoecke2009,Coecke2009Picturalism,chiribella2010probabilistic,Hardy_2012} helped to set the stage for the development of quantum causal models.  Although they were conceived independently of the framework of classical causal models, they were quite similar to that framework insofar as they made heavy use of DAGs---in the form of circuit diagrams---to depict structural features of a set of processes. When the authors of these formulations turned their attention to relativistic causal structure, the frameworks they devised drew even closer in spirit to that of causal models. Prominent examples include: the causaloid framework of Hardy~\cite{Hardy_2007}, the multi-time formalism of Aharonov and co-workers~\cite{aharonov2007two,AharonovEtAl_2009,aharonov2013each,silva2014pre}, the quantum combs framework of Chiribella, D'Ariano and Perinotti~\cite{ChiribellaEtAl_2009,Chiribella_2012,ChiribellaEtAl_2013}, the causal categories of Coecke and Lal~\cite{coecke2013causal}, and the process matrix formalism of Oreshkov, Costa and Brukner~\cite{Oreshkov2012,AraujoEtAl_2014}.  
A common aim of these approaches is to be able to compute the consequences of an intervention upon a particular quantum system within the circuit, and this is precisely one of the tasks that a quantum analogue of a causal model should be able to handle.

Many of these frameworks 
represent a quantum system at a given region of space-time by two copies of its Hilbert space, one corresponding to the system that is input into the region and one corresponding to the system that is output from it.  In this way, the region becomes a ``locus of intervention'' for the system.  By inserting a particular quantum process into the ``slot'', one determines the nature of the intervention. 
This is the approach taken, for instance, in the multi-time formalism of Ref.~\cite{AharonovEtAl_2009}, the quantum combs of Ref.~\cite{ChiribellaEtAl_2009}, and the process matrices of Ref.~\cite{Oreshkov2012}.   This representation of interventions has a counterpart in classical causal models, for instance in the work of \cite{richardson2013single}, as was noted in Refs.~\cite{Ried-Agnew-15,Costa2016}. 

Costa and Shrapnel~\cite{Costa2016} in particular have sought to explicitly cast this sort of framework as a quantum generalization of a causal model.  In their approach, the nodes of the DAG are associated with a quantum system localized in a region (understood as a potential locus of intervention) and the collection of edges from one set of nodes to another represent causal processes.  

An approach of this sort is required  if one seeks to find intrinsically quantum versions of important theorems of classical causal models. 
For instance, while Henson, Lal and Pusey~\cite{Henson-Lal-14} derive a generalization of the d-separation theorem of classical causal models, it only 
infers conditional independence relations from d-separation relations for the classical variables in the graph.
 An intrinsically quantum version of the d-separation theorem, by contrast, would be one which concerns the causal relations among {\em quantum systems} (see, for instance, \cite{Pienaar-Brukner-15}).  If a set of nodes representing quantum systems can be described by a joint or conditional state, then one can seek to determine whether factorization conditions on this state are implied by d-separation relations among the quantum systems on the graph.   
Similarly, while both the approaches of Henson, Lal and Pusey~\cite{Henson-Lal-14} and of Fritz~\cite{fritz-14} allow one to derive, from the structure of the DAG, constraints on the joint distribution over classical variables embedded therein, they do not address an intrinsically quantum version of this problem.  If a set of nodes representing quantum systems can be described by a joint or conditional state, then one can seek to derive constraints on this state directly from the structure of the DAG.

Our own approach aims at an intrinsically quantum generalization of the notion of a causal model. We therefore associate to each node of the DAG a quantum system localized to a space-time region, and we represent it by a pair of Hilbert spaces, corresponding to the input and output of an intervntion upon the system. 
Consequently, our approach is very similar to that of Costa and Shrapnel~\cite{Costa2016}.  Nonetheless, there is a significant difference in how we represent common causes. 

In Costa and Shrapnel's work, any node with multiple outgoing edges is represented as a locus of intervention where the output Hilbert space is a tensor product of Hilbert spaces, one for each outgoing edge.  As such, any node acting as a common cause must be associated with a composite quantum system.  It cannot, for instance, be associated with a single qubit.  
By contrast, our approach does not constrain the representation of common causes in this fashion.  Any quantum system, including a single qubit, may constitute a complete common cause of a collection of other quantum systems.  This extra generality is required since, as our examples have shown, the complete common cause of a set of systems {\em can} be a single qubit.  Second, and more importantly, our work has shown that for a quantum channel whose input is the complete common cause of its $n$ outputs, it is not the case that the channel must split the input into $n$ components, each of which exerts a causal influence on a different output. 
This is merely one special case of the most general form that such a channel can take.  Third, if the complete common causes consist of multiple nodes in the DAG, then it is only the joint Hilbert space of the collection of these that must satisfy the condition of factorizing-in-subspaces, while each individual Hilbert space need not. 

These differences are likely to have a significant impact on the form of any intrinsically quantum d-separation theorem.

Finally, we mention a third purpose to which quantum causal models can be put.
Theorems about classical causal models often concern the sorts of {\em inferences} one can make about one variable given information about another.  As an example, if $Z$ is a common effect of $X$ and $Y$, then learning $Z$ can induce correlations between $X$ and $Y$.  As such, one might expect quantum causal models to also constrain the sort of inferences one can make among quantum variables.  Early work by Leifer and Spekkens~\cite{Leifer-Spekkens-13} had this purpose in mind.  The authors noted various scenarios in which their proposal could not be applied, and subsequent work~\cite{horsman2016can} has narrowed down the scope of possibilities for any such proposal.  Our own proposal provides the means of making many of the Bayesian inferences considered in Ref.~\cite{Leifer-Spekkens-13}.  The case discussed in Sec.~\ref{bayesianupdating} is one such example.

There is also prior work on quantum causal models that takes a significantly different approach to the ones described above and for which the relation to our work is less clear.  The work of Tucci~\cite{tucci1995quantum,tucci2012introduction}, which is in fact the earliest attempt at a quantum generalization of a causal model, represents causal dependences by complex transition amplitudes rather than quantum channels.

\section{Conclusions}\label{conclusions}

The field of classical statistics has benefited greatly from analysis provided by the formalism of causal models~\cite{Pearl-09, Spirtes-Glymour-01}. In particular, this formalism allows one to infer facts about the underlying causal structure purely from uncontrolled statistical data, a tool with significant applications in all branches of the physical and social sciences. Given that some seemingly paradoxical features of classical correlations have found satisfying resolutions when viewed through a causal lens, one might wonder to what extent the same is true of quantum correlations.

Starting with the idea that whatever innovation quantum theory might hold for causal models, the intuition contained in Reichenbach's principle ought to be preserved, we motivated the problem of finding a quantum version of the principle.  This required us to determine what constraint a channel from $A$ to $BC$ must satisfy if $A$ is the complete common cause of $B$ and $C$.  We solved the problem by considering a unitary dilation of the channel and by noting that there is no ambiguity in how 
to represent the absence of causal influences between certain inputs and certain outputs of a unitary.  From this, we derived a notion of quantum conditional independence for the outputs of the channel given its input.  This inference from a common-cause structure to quantum conditional independence was then generalized to obtain our quantum version of causal models.

Given a state on a quantum causal model, we described how to construct a marginal state for a subset of nodes.  We discussed a number of simple examples of quantum channels and causal structures. A theme of the examples is that when there is a difference between the quantum and classical cases, this can often be understood if one takes the view that a quantum pure state represents maximal but incomplete information about a system, and hence may underwrite correlations between other systems in a way that a classical pure state cannot.

There are many directions for future work. In the case of classical causal networks, an important theorem states that the \emph{d-separation} relation among nodes of a DAG is sound and complete for a conditional independence relation to hold among the associated variables in the joint probability distribution \cite{Pearl-09}. Here, for arbitrary subsets of nodes $S$, $T$ and $U$, subsets $S$ and $T$ are said to be d-separated by $U$ if a certain criterion holds, where this is determined purely by the structure of the DAG. An important question is therefore whether d-separation is sound and complete for some natural property of the state $\sigma$ on a quantum causal network. 

It is also desirable to relate properties of a quantum causal network to operational statements involving the outcomes of agents' interventions: under what circumstances, for example, does it follow that there is an intervention by the agents at nodes in a subset $U$, such that, conditioned on its outcome, the outcomes of any interventions by agents at $S$ and $T$ must be independent? Such a result would have an application to quantum protocols. Imagine, for example, a cryptographic scenario in which agents at $S$ and $T$ desire shared correlations that are not screened off by the information held by agents at $U$. 

In the classical case, there has been a great deal of work on the problem of \emph{causal inference} \cite{Pearl-09,lee2015causal,wolfe2016inflation,chaves2014inferring}: given only certain facts about the joint probabilities, for instance, a set of conditional independences, what can be inferred about the underlying causal structure? For an initial approach to quantum causal inference, with a quantum-over-classical advantage in a simple scenario, see \cite{Ried-Agnew-15}. The formalism of quantum causal networks described here is the appropriate framework for inferring facts about underlying, intrinsically quantum, causal structure, given observed facts about the outcomes of interventions by agents. 

Recently, there has been much interest in deriving bounds on the correlations achievable in classical causal models \cite{lee2015causal, wolfe2016inflation, chaves2016polynomial, rosset2016nonlinear} using insights from the literature on Bell's theorem. Such bounds constitute Bell-like inequalities for arbitrary causal structures. 
The main technical challenge in deriving such inequalities is that the set of correlations is generally not convex if the DAG has more than one latent variable, so that standard techniques for deriving Bell inequalities are not applicable. 
By adapting these new techniques to the formalism presented here, one could perhaps systematically derive bounds on the quantum correlations achievable in certain quantum causal models thereby providing a general method of deriving Tsirelson-like bounds for arbitrary causal structures.

Finally,  it would be interesting to extend the formalism to explore the possibility that certain quantum scenarios are best understood as involving a quantum-coherent combination of different causal structures~ \cite{Chiribella_2012,Oreshkov2012,maclean2016quantum,feix2016quantum}.  It has been argued that the possibility of such indefinite causal structure may be significant for the project of unifying quantum theory with general relativity~\cite{Hardy_2007}.

\begin{acknowledgments}
RWS thanks Elie Wolfe for helpful discussions. This work was supported by EPSRC grants, the EPSRC National Quantum Technology Hub in Networked Quantum Information Technologies, an FQXi Large Grant, University College Oxford, the Wiener-Anspach Foundation, and by the Perimeter Institute for Theoretical Physics. Research at Perimeter Institute is supported by the Government of Canada through the Department of Innovation, Science and Economic Development Canada and by the Province of Ontario through the Ministry of Research, Innovation and Science. This project/publication was made possible through the support of a grant from the John Templeton Foundation. The opinions expressed in this publication are those of the author(s) and do not necessarily reflect the views of the John Templeton Foundation.
\end{acknowledgments}

\bibliography{bibliography}

\onecolumngrid
\appendix

\section{Proof of Theorems~\ref{maintheorem} and \ref{alternativeexpressions}}\label{maintheoremproof}

We here provide the proof of Thms~\ref{maintheorem} and \ref{alternativeexpressions}.  This amounts to proving that for a channel $\rho_{BC|A}$, the following four conditions are equivalent:
\begin{enumerate}
\item $\rho_{BC|A}$ admits of a unitary dilation where $A$ is a complete common cause of $B$ and $C$. 
\item $\rho_{BC|A} = \rho_{B|A} \rho_{C|A}$.
\item $I(B:C|A) = 0$ where  $I(B:C|A)$ is the quantum conditional mutual information evaluated on the (positive, trace-one) operator $\hat{\rho}_{BC|A}$.
\item The Hilbert space for the $A$ system can be decomposed as $\mathcal{H}_{A} = \bigoplus_i \mathcal{H}_{A_i^L}\otimes\mathcal{H}_{A_i^R}$ and $\rho_{BC|A} = \sum_i \left(\rho_{B|A_i^L}\otimes\rho_{C|A_i^R}\right)$, where for each $i$, $\rho_{B|A_i^L}$ represents a completely positive map $\mathcal{B}(\mathcal{H}_{A_i^L}) \rightarrow \mathcal{B}(\mathcal{H}_B)$, and $\rho_{C|A_i^R}$ a completely positive map $\mathcal{B}(\mathcal{H}_{A_i^R}) \rightarrow \mathcal{B}(\mathcal{H}_C)$.
\end{enumerate}

We will show various implications that collectively give Thm~\ref{maintheorem}.
\vskip15pt
{\bf Proof that $(3) \leftrightarrow (4)$.}
\vskip15pt
This follows easily from the results of Ref.~\citep{Hayden-Jozsa-04}, where a characterization is given of tripartite quantum states over systems $A,B,C$ that satisfy $I(B:C|A)=0$.
\begin{lemma}[\cite{Hayden-Jozsa-04} Thm~6] \label{thm:hayden-jozsa}
For any tripartite quantum state $\rho_{ABC}$, the quantum conditional mutual information $I(B:C|A)=0$ if and only if the Hilbert space of the $A$ system decomposes as $\mathcal{H}_{A} = \bigoplus_i \mathcal{H}_{A_i^L}\otimes\mathcal{H}_{A_i^R}$, such that
\begin{equation}\label{haydendecomp}
\rho_{ABC} = \sum_i p_i \left(\rho_{BA_i^L}\otimes\rho_{CA_i^R}\right),\quad p_i\geq0,\quad\sum_i p_i=1,
\end{equation}
where for each $i$, $\rho_{BA_i^L}$ is a quantum state on $\mathcal{H}_B\otimes \mathcal{H}_{A_i^L}$ and  $\rho_{CA_i^R}$ is a quantum state on $\mathcal{H}_C\otimes \mathcal{H}_{A_i^R}$.
\end{lemma}
Thm~\ref{maintheorem} concerns the channel operator $\rho_{BC|A}$, which satisfies $\mathrm{Tr}_{BC}(\rho_{BC|A}) = I_{A^\ast}$. Applying Lem.~\ref{thm:hayden-jozsa} to the operator $\hat{\rho}_{BC|A} = (1/d_A) \rho_{BC|A}$ yields the decomposition
\[
\hat{\rho}_{BC|A} = \sum_i p_i \left(\hat{\rho}_{B|A_i^L}\otimes\hat{\rho}_{C|A_i^R}\right).
\]
Using $\mathrm{Tr}_{BC}(\hat{\rho}_{BC|A}) = (1/d_A) I_{A^\ast}$, it follows that for each $i$, the components satisfy $\mathrm{Tr}_B (\hat{\rho}_{B|A_i^L}) = (1/d_{A_i^L}) I_{(A_i^L)^\ast}$, and $\mathrm{Tr}_C (\hat{\rho}_{C|A_i^R}) = (1/d_{A_i^R}) I_{(A_i^R)^\ast}$, with $p_i = (d_{A_i^L} d_{A_i^R})/d_{A}$. The result follows.

\vskip15pt
{\bf Proof that $(1) \rightarrow (4)$.}
\vskip15pt

Let $\rho^U_{BFC|A\lambda_B\lambda_C}$ be the Choi-Jamio\l{}kowski operator for the unitary $U$, defined according to the conventions set out in the main text. Let missing indices indicate that a partial trace is taken, as also in the main text.
Note that in general $\rho^U_{BC|A}\neq\rho_{BC|A}$, since the latter is obtained via a particular choice of input states for $\lambda_B$ and $\lambda_C$. The proof proceeds by proving relations between quantum conditional mutual informations evaluated on the renormalized operator $\hat{\rho}^U_{BFC|A\lambda_B\lambda_C} = (1/d_{\lambda_B}d_{A}d_{\lambda_C}) \rho^U_{BFC|A\lambda_B\lambda_C}$, and its partial traces. 

First,
\begin{equation} \label{eqn:lem13-firstentopyrelation}
I(B:FC | \lambda_B A \lambda_C )=0.
\end{equation}
This follows by expanding in terms of von Neumann entropies:
\begin{equation}
I(B:FC | \lambda_B A \lambda_C ) = S(\hat{\rho}^U_{B| \lambda_B A \lambda_C  }) + S(\hat{\rho}^U_{FC| \lambda_B A \lambda_C }) - S(\hat{\rho}^U_{ BFC | \lambda_B A \lambda_C}) - S(\hat{\rho}^U_{\cdot |  \lambda_B A \lambda_C  }).
\end{equation}
The third term is zero, since the unitarity of $U$ implies that $\hat{\rho}^U_{BFC|\lambda_B A \lambda_C}$ is a pure state. The final term is $\log(d_{\lambda_B} d_{A} d_{\lambda_C})$, since $\hat{\rho}^U_{ \cdot | \lambda_B A \lambda_C} = (1/d_{\lambda_B} d_{A} d_{\lambda_C}) I_{(\lambda_B A \lambda_C)^\ast}$. Noting also that 
$\mathrm{Tr}_{\lambda_B A \lambda_C} (\hat{\rho}^U_{ BFC | \lambda_B A \lambda_C}) = (1/d_{\lambda_B} d_{A} d_{\lambda_C}) I_{(\lambda_B A \lambda_C)^\ast}$, and using the fact that the von Neumann entropy of the partial trace of a pure state is equal to the von Neumann entropy of the complementary partial trace, yields that the first two terms equal $\log (d_F d_C)$ and $\log (d_B)$ respectively, hence their sum is equal to $\log(d_{\lambda_B} d_{A} d_{\lambda_C})$, and Eq.~(\ref{eqn:lem13-firstentopyrelation}) follows.

Second, 
\begin{equation} \label{eqn:lem13-secondentropyrelation}
I(\lambda_B : \lambda_C | A )=0.
\end{equation}
This follows immediately from $\hat{\rho}^U_{ \cdot | \lambda_B A \lambda_C} = (1/d_{\lambda_B} d_{A} d_{\lambda_C}) I_{(\lambda_B A \lambda_C)^\ast}$.

Third,
\begin{equation} \label{eqn:lem13-thirdentropyrelation}
I(B : \lambda_C | \lambda_B A) = 0.
\end{equation}
To see this, write
\begin{equation}
I(B : \lambda_C | \lambda_B A)  = S(\hat{\rho}^U_{B | \lambda_B A}) + S(\hat{\rho}^U_{\cdot | \lambda_B A \lambda_C}) - S(\hat{\rho}^U_{B | \lambda_B A \lambda_C}) - S(\hat{\rho}^U_{\cdot | \lambda_B A}).
\end{equation}
The second and fourth terms are entropies of maximally mixed states on their respective systems, hence sum to $\log d_{\lambda_C}$. For the first and third terms, it follows from the assumption that there is no causal influence from $\lambda_C$ to $B$ in $U$ that $\hat{\rho}^U_{B | \lambda_B A \lambda_C} = \hat{\rho}^U_{B | \lambda_B A} \otimes (1/d_{\lambda_C}) I_{(\lambda_C)^\ast}$. Hence the third term is equal to $S(\hat{\rho}^U_{B | \lambda_B A}) + \log (d_{\lambda_C})$, which gives Eq.~(\ref{eqn:lem13-thirdentropyrelation}).

Fourth,
\begin{equation} \label{eqn:lem13-fourthentropyrelaiton}
I(C : \lambda_B | A \lambda_C) = 0.
\end{equation}
This follows from a similar argument as Eq.~(\ref{eqn:lem13-thirdentropyrelation}), using the assumption that there is no influence from $\lambda_B$ to $C$ in $U$. 

The aim is now to use Eqs.~(\ref{eqn:lem13-firstentopyrelation},\ref{eqn:lem13-secondentropyrelation},\ref{eqn:lem13-thirdentropyrelation},\ref{eqn:lem13-fourthentropyrelaiton}) to show that $\hat{\rho}_{BC|A}$ satisfies $I(B:C|A)=0$. This follows using a result from Ref.~\citep{Leifer-Poulin-08}, which states that quantum conditional mutual informations on partial traces of a multipartite quantum state satisfy the \emph{semi-graphoid axioms} familiar from the classical formalism of causal networks \cite{Pearl-09}. The semi-graphoid axioms are as follows:
\begin{align} \label{eqn:lem13-semi-graphoid1}
\left[I(X:Y|Z) = 0\right]  &\Rightarrow \left[I(Y:X|Z) = 0\right] \\
\left[I(X:YW|Z) = 0\right] &\Rightarrow \left[I(X:Y|Z) = 0\right] \\
\left[I(X:YW|Z) = 0\right] &\Rightarrow \left[I(X:Y|ZW) = 0\right] \\
\left[I(X:Y|Z) = 0 \right]
\wedge \left[I(X:W|YZ) = 0\right] &\Rightarrow \left[I(X:YW|Z) = 0\right] \label{eqn:lem13-semi-graphoid-4}
\end{align}

Applying Eqs.~(\ref{eqn:lem13-semi-graphoid1}-\ref{eqn:lem13-semi-graphoid-4}) to Eqs.~(\ref{eqn:lem13-firstentopyrelation},\ref{eqn:lem13-secondentropyrelation},\ref{eqn:lem13-thirdentropyrelation},\ref{eqn:lem13-fourthentropyrelaiton}) gives
\begin{align}
\left[I(B  : FC    | \lambda_B A \lambda_C ) = 0 \right] &\Rightarrow \left[I(B : C | \lambda_B A \lambda_C) = 0\right] \\
\left[I(C: \lambda_B | A\lambda_C) = 0 \right]
\wedge \left[I(B : C | \lambda_B A \lambda_C) = 0\right] &\Rightarrow \left[I(C : B\lambda_B | A\lambda_C) = 0\right] \\
\left[I(\lambda_B : \lambda_C | A) = 0 \right]
\wedge \left[I(\lambda_C : B | \lambda_B A) = 0\right] &\Rightarrow \left[I(\lambda_C : B \lambda_B | A) = 0\right] \\
\left[I(B \lambda_B : \lambda_C | A) = 0\right]
\wedge \left[I(B \lambda_B : C | A \lambda_C) = 0\right] &\Rightarrow \left[I(B\lambda_B : C \lambda_C | A) = 0\right]
\end{align}

Hence condition~(1) of the theorem implies that $I(B\lambda_B : C \lambda_C | A) = 0$, where this quantity is calculated on the trace-one Choi-Jamio\l{}kowski operator representing the dilation unitary $U$. Using Lem.~\ref{thm:hayden-jozsa} gives
\begin{equation} 
\hat{\rho}^U_{BC | \lambda_B A \lambda_C} = \sum_i p_i \left( \hat{\rho}^U_{B|\lambda_B A_i^L} \otimes \hat{\rho}^U_{C|A_i^R \lambda_C} \right),
\end{equation}
for some appropriate decomposition of $(\mathcal{H}_A)^\ast$ and probability distribution $\{p_i\}_i$. The form of the decomposition, and the fact that $\mathrm{Tr}_{BC}(\hat{\rho}^U_{BC | \lambda_B A \lambda_C}) = (1/d_{\lambda_B} d_{A} d_{\lambda_C}) I_{(\lambda_B A \lambda_C)^\ast}$, gives
\begin{equation}
\rho^U_{BC | \lambda_B A \lambda_C} = \sum_i \left( \rho^U_{B|\lambda_B A_i^L} \otimes \rho^U_{C|A_i^R \lambda_C} \right),
\end{equation}
where for each $i$, the components satisfy $\mathrm{Tr}_B(\rho^U_{B|\lambda_B A_i^L}) = I_{(A_i^L)^\ast}$ and $\mathrm{Tr}_C(\rho^U_{C|\lambda_C A_i^R}) = I_{(A_i^R)^\ast}$. The operator $\rho_{BC|A}$ is obtained by acting with this channel on the input states $|0\rangle_{\lambda_B}$ for $\lambda_B$ and $|0\rangle_{\lambda_C}$ for $\lambda_C$. This gives 
\[
\rho_{BC | A} = \sum_i \left( \rho_{B| A_i^L} \otimes \rho_{C|A_i^R } \right),
\]
where $\mathrm{Tr}_B(\rho_{B | A_i^L}) = I_{(A_i^L)^\ast}$ and $\mathrm{Tr}_C(\rho_{C | A_i^R}) = I_{(A_i^R)^\ast}$, as required.

\vskip15pt
{\bf Proof that $(4) \rightarrow (1)$.}
\vskip15pt

Let $\mathcal{H}_{A} = \bigoplus_i \mathcal{H}_{A_i}$, with $\mathcal{H}_{A_i} = \mathcal{H}_{A_i^L}\otimes\mathcal{H}_{A_i^R}$, and $\rho_{BC|A} = \sum_i \left(\rho_{B|A_i^L}\otimes\rho_{C|A_i^R}\right)$. Each term $\rho_{B|A_i^L}$ corresponds to a valid quantum channel, i.e., a CPTP map $\mathcal{B}(\mathcal{H}_{A_i^L}) \rightarrow \mathcal{B}(\mathcal{H}_B)$. Similarly, each term $\rho_{C|A_i^R}$ corresponds to a CPTP map $\mathcal{B}(\mathcal{H}_{A_i^R}) \rightarrow \mathcal{B}(\mathcal{H}_C)$. 

The channel $\rho_{B|A_i^L}$ can be dilated to a unitary transformation $V_i$, with ancilla input $\lambda_B$ in a fixed state $|0\rangle_{\lambda_B}$, such that $V_i$ acts on the Hilbert space $\mathcal{H}_{\lambda_B}\otimes \mathcal{H}_{A_i^L}$. Similarly, $\rho_{C|A_i^R}$ can be dilated to a unitary transformation $W_i$, with ancilla $\lambda_C$ in a fixed state $|0\rangle_{\lambda_C}$, acting on $\mathcal{H}_{A_i^R}\otimes \mathcal{H}_{\lambda_C}$. By choosing the dimension of $\lambda_B$ large enough, we can identify the system $\lambda_B$ and the state $|0\rangle_{\lambda_B}$ that are used for each value of $i$, and similarly $\lambda_C$.

Let $V_i'$ be the operator that acts as $V_i\otimes I_{A_i^R} $ on the subspace $\mathcal{H}_{\lambda_B} \otimes \mathcal{H}_{A_i}$, and as zero on the subspace $\mathcal{H}_{\lambda_B} \otimes \mathcal{H}_{A_j}$ for $j\ne i$. 
Similarly, let $W_i'$ be the operator that acts as $I_{A_i^L}\otimes W_i$ on the subspace $\mathcal{H}_{A_i}\otimes \mathcal{H}_{\lambda_C}$, and as zero on the subspace $\mathcal{H}_{A_j}\otimes \mathcal{H}_{\lambda_C}$ for $j\ne i$. Let
\begin{align}
V &= \sum_i V_i' \\
W &= \sum_i W_i',
\end{align}
where $W$ and $V$ are unitary and $[V\otimes I_{\lambda_C} , I_{\lambda_B} \otimes W] = 0$. 
The channel represented by $\rho_{BC|A}$ can be dilated to the unitary transformation $U = (I_{\lambda_B} \otimes W)(V\otimes I_{\lambda_C})$, with ancillas $\lambda_B$ and $\lambda_C$. From the form of $V$ and $W$, it follows immediately that there is no causal influence from $\lambda_C$ to $B$ in $U$. From $[V\otimes I_{\lambda_C} , I_{\lambda_B} \otimes W] = 0$ and the form of $V$ and $W$, it follows immediately that there is no influence from $\lambda_B$ to $C$ in $U$.

\vskip15pt
{\bf Proof that $(2) \rightarrow (3)$.}
\vskip15pt

As remarked in the main text, taking the Hermitian conjugate of $\rho_{BC|A}=\rho_{B|A}\rho_{C|A}$ immediately gives $[\rho_{B|A},\rho_{C|A}]=0$. Hence
\begin{align}
\rho_{BC|A}  = & \rho_{B|A}\rho_{C|A} \\
\rho_{BC|A}  = & \exp\left[ \log\rho_{B|A} + \log\rho_{C|A} \right] \\
\log\rho_{BC|A}  = & \log\rho_{B|A} + \log\rho_{C|A} \\
\log\rho_{BC|A} + \log\rho_{\cdot|A}  = & \log\rho_{B|A} + \log\rho_{C|A} \\
\log(d_A^{-1}\rho_{BC|A}) + \log(d_A^{-1}\rho_{\cdot|A})  = & \log(d_A^{-1}\rho_{B|A}) + \log(d_A^{-1}\rho_{C|A})
\end{align}
The second line follows as $[\rho_{B|A},\rho_{C|A}]=0$; the fourth because $\rho_{\cdot|A}=I_{A^\ast}$ and therefore has the zero matrix as its logarithm; and the final line by adding $2\log d_A^{-1}$ to both sides. It is proved in Ref.~\cite{ruskai2002inequalities} that for any trace-one density operator $\rho_{XYZ}$ then $\log\rho_{XYZ} + \log\rho_{Z}  =  \log\rho_{XZ} + \log\rho_{YZ}$ is equivalent to the condition $I(X:Y|Z)=0$.

\vskip15pt
{\bf Proof that $(4) \rightarrow (2)$.}
\vskip15pt

Condition (4) is that $\mathcal{H}_{A} = \bigoplus_i \mathcal{H}_{A_i^L}\otimes\mathcal{H}_{A_i^R}$, with $\rho_{BC|A} = \sum_i \left(\rho_{B|A_i^L}\otimes\rho_{C|A_i^R}\right)$. It follows that 
\begin{align}
\rho_{B|A} & =  \sum_{i} \left( \rho_{B|A_{i}^{L}}\otimes I_{(A_{i}^{R})^\ast} \right), \\
\rho_{C|A} & =  \sum_{i} \left( I_{(A_{i}^{L})^\ast}\otimes\rho_{C|A_{i}^{R}} \right).
\end{align}
The product is
\begin{equation}
\rho_{B|A}\rho_{C|A} = \sum_{i,j} \left( \rho_{B|A_{i}^{L}}\otimes I_{(A_{i}^{R})^\ast} \right) \left( I_{(A_{j}^{L})^\ast}\otimes\rho_{C|A_{j}^{R}} \right).
\end{equation}
The only non-zero terms correspond to  $i=j$, hence
\begin{equation}
\rho_{B|A}\rho_{C|A} = \sum_{i} \rho_{B|A_{i}^{L}}\otimes\rho_{C|A_{i}^{R}} = \rho_{BC|A}.
\end{equation}

\section{Proof of Theorem~\ref{maintheoremmany}}\label{maintheoremmanyproof}

\vskip15pt
{\bf Proof that $(3)\rightarrow(2)$.}
\vskip15pt

The proof proceeds via an inductive argument. Consider  
$$\rho_{B_1\dots B_n|A}=\mathrm{Tr}_{B_{n+1}\dots B_k}\big(\rho_{B_1\dots B_k|A}\big),$$
with $2 \leq n<k$, and assume that the claim holds for this channel, hence
 \begin{equation} \label{induction}
 \rho_{B_1\dots B_n|A}=\rho_{B_1|A}\cdots\rho_{B_n|A},
 \end{equation}
with $[\rho_{B_i|A},\rho_{B_j|A}]=0$ for $i,j=1,\dots,n$. It will be shown that the claim remains true if one fewer system is traced out. To see this, recall that condition $(3)$ gives $I(B_{n+1}:\overline{B}_{n+1}|A)=0$, where $I(B_{n+1}:\overline{B}_{n+1}|A)$ is evaluated on $\hat{\rho}_{B_1\dots B_k|A}$. Using Thm~\ref{maintheorem} gives
$$\rho_{B_1\dots B_{k}|A}=\rho_{\overline{B}_{n+1}|A}\rho_{B_{n+1}|A},$$
with $[\rho_{\overline{B}_{n+1}|A},\rho_{B_{k+1}|A}]=0.$ Tracing out systems $B_{n+2}\dots B_{k}$ results in
$$\rho_{B_1\dots B_{n+1}|A}=\rho_{B_1\dots B_{n}|A}\rho_{B_{n+1}|A},$$
with $[\rho_{B_1\dots B_{n}|A},\rho_{B_{n+1}|A}]=0.$ Since $\rho_{B_1\dots B_{n}|A}$ satisfies Eq.~(\ref{induction}) it follows that
$$
 \rho_{B_1\dots B_{n+1}|A}=\rho_{B_1|A}\cdots\rho_{B_{n+1}|A}.
 $$
For any $i=1,\dots, n$, trace out all systems but $B_i,B_{n+1}$ and $A$ to see that $[\rho_{B_i|A},\rho_{B_{n+1}|A}]=0$.
 
Hence if $\rho_{B_1\dots B_{n}|A}$ satisfies the claim, so too does $\rho_{B_1\dots B_{n+1}|A}.$ As $\rho_{B_1B_2|A}=\rho_{B_1|A}\rho_{B_2|A},$ with $[\rho_{B_1|A},\rho_{B_{2}|A}]=0$ follows from $I(B_{1}:\overline{B}_{1}|A)=0$, and tracing out all but systems $B_1,B_2,$ and $A$, the proof is complete.

\vskip15pt
{\bf Proof that $(2)\rightarrow(3)$.}
\vskip15pt

This is immediate from Thm~\ref{maintheorem}, by grouping outputs into $B_i$ and $\overline{B}_i$ for each $i$.

\vskip15pt
{\bf Proof that $(3) \leftrightarrow (4)$.}
\vskip15pt 

The proof that $(4)\rightarrow (3)$ is immediate from Thm~\ref{maintheorem}, by grouping outputs into $B_i$ and $\overline{B}_i$ for each $i$.

It remains to show that if $I(B_i:\overline{B}_i|A)=0 \text{ for all } i,$ then there exists a decomposition 
\begin{equation} \label{multi decomposition}
\mathcal{H}_A=\bigoplus_i\left(\bigotimes_{j=1}^k\mathcal{H}_{A_i^j}\right),
\end{equation}  
with $\rho_{B_1\ldots B_k|A} = \sum_i ( \rho_{B_1 | A^1_i} \otimes\cdots\otimes \rho_{B_k | A^k_i})$.

Given $I({B}_1:\overline{B}_1|A)=0$, Thm~\ref{maintheorem} implies that $\mathcal{H}_{A}$ decomposes as
$$\mathcal{H}_{A} = \bigoplus_i \mathcal{H}_{A_i^L}\otimes\mathcal{H}_{A_i^R},$$
with $\rho_{B_1\dots B_k|A}=\sum_{i}\rho_{B_1|A_{i}^{L}}\otimes\rho_{{B}_2\dots B_k|A_{i}^{R}}.$ By assumption, $I(B_2:\overline{B}_2|A)=I(B_2:B_1,B_3,\dots,B_k|A)=0$. As the conditional mutual information never increases if systems are discarded, we have $0=I(B_2:B_1,B_3,\dots,B_k|A)\geq I(B_2:B_3,\dots,B_k|A)$. Non-negativity of the conditional mutual information then yields $ I(B_2:B_3,\dots,B_k|A)=0$. 

The above decomposition ensures 
$$\hat{\rho}_{B_2\dots B_k|A} = \sum_i p_i \left( \frac{I_{A^L_i}}{d_{A_i^L}} \right) \otimes \hat{\rho}_{{B}_2\dots B_k|A_{i}^{R}},$$
with $p_i = d_{A_i^L} d_{A_i^R} / d_A$. As the terms in the sum on the RHS have support on orthogonal subspaces, 
$$\begin{aligned} 
S(\hat{\rho}_{B_2B_3\dots B_k|A}) &= H(p) + \sum_i p_i \log d_{A_i^L} + \sum_ip_iS(\hat{\rho}_{{B}_2\dots B_k|A_{i}^{R}}), \\
S(\hat{\rho}_{B_2|A})&=H(p)+\sum_i p_i \log d_{A_i^L}+ \sum_ip_iS(\hat{\rho}_{{B}_2|A_{i}^{R}}), \\
S(\hat{\rho}_{B_3\dots B_k|A})&=H(p)+\sum_i p_i \log d_{A_i^L}+\sum_ip_iS(\hat{\rho}_{{B}_3\dots B_k|A_{i}^{R}}), \\
S(\hat{\rho}_{\cdot |A})&=H(p)+\sum_i p_i \log d_{A_i^L}+\sum_ip_iS(\hat{\rho}_{\cdot |A_{i}^{R}}). 
\end{aligned}$$
Substituting into
$$I(B_2:B_3,\dots,B_k|A)= S(\hat{\rho}_{B_2|A}) + S(\hat{\rho}_{B_3\dots B_k|A}) - S(\hat{\rho}_{B_2B_3\dots B_k|A}) - S(\hat{\rho}_{\cdot |A}),$$
the $H(p)$ terms and the $\sum_i p_i \log d_{A_i^L}$ terms cancel, and one is left with 
$$I(B_2:B_3,\dots,B_k|A)=\sum_i p_i I(B_2:B_3,\dots,B_k|A_i^R)=0.$$
Non-negativity of both the conditional mutual information and the $p_i$ implies $$I(B_2:B_3\dots B_k|A_i^R)=0.$$ Hence each $\mathcal{H}_{A_i^R}$ in the above decomposition further decomposes into a direct sum of tensor products. Iterating this procedure results in the required decomposition.

\vskip15pt
{\bf Proof that $(1) \leftrightarrow (4)$.}
\vskip15pt   

The proof that $(4)\rightarrow (1)$ is a straightforward extension of the proof in Appendix~\ref{maintheoremproof} that condition $(4)\rightarrow (1)$ in Thm~\ref{maintheorem}. 

To show that $(1)\rightarrow (4)$, first use Def.~\ref{nonsig} to show that if, for each $i$, there is no causal influence from $\lambda_i$ to $\overline{B}_i$, it follows that, for each $i$, there is no causal influence from $\overline{\lambda}_i$ to $B_i$. Partitioning the output systems into $B_i$ and $\overline{B}_i$, and the ancilla systems $\lambda_1,\dots, \lambda_k$ into $\lambda_i$ and $\overline{\lambda}_i$, it follows from Thm~\ref{maintheorem} that $I(B_i:\overline{B}_i|A)=0$. Hence condition $(1) \rightarrow (3)$, and since condition $(3) \rightarrow (4)$, the result follows.

\end{document}